\documentclass[aps,reprint]{revtex4-2}
\usepackage{etoolbox} 
\usepackage{blindtext}
\usepackage{graphicx}
\usepackage{subcaption}
\usepackage{amssymb}
\usepackage{xcolor}
\usepackage{graphicx}
\usepackage{floatrow}
\usepackage[fleqn]{amsmath}
\usepackage[font=small,skip=3pt]{caption}
\usepackage{makecell}
\usepackage{hyperref} 
\usepackage{cleveref}
\usepackage{url}
\usepackage{siunitx}
\usepackage{float}
\usepackage{braket}
\usepackage{bm}

\makeatletter
\appto{\appendix}{%
  \@ifstar{\def\theequation@prefix{A.}}%
          {}
}
\makeatother

\newcolumntype{B}{>{\color{red}}c}

\begin{document}
\title{Persistence of Deuterium and Tritium Nuclear Spin-Polarization in Presence of High-Frequency Plasma Waves}
\author{J. W. S. Cook$^{1,2,3}$}
\author{H. Ali$^{4}$}
\author{J. F. Parisi$^{4}$}
\author{A. Diallo$^{5}$}
\author{N. Faatz$^{6,7,8}$}
\affiliation{$^1$Fourth State Labs Ltd, Culham Campus, United Kingdom, OX14 3DB, UK}
\affiliation{$^2$United Kingdom Atomic Energy Authority, Culham Campus, Abingdon, Oxfordshire, OX14 3DB, United Kingdom}
\affiliation{$^3$Centre for Fusion, Space and Astrophysics, Department of Physics, University of Warwick, Coventry, CV4 7AL, UK}
\affiliation{$^4$Marathon Fusion, 150 Mississippi, San Francisco, 94107, CA, USA}
\affiliation{$^5$Princeton Plasma Physics Laboratory, Princeton University, Princeton, NJ, USA}
\affiliation{$^6$GSI, Helmholtzzentrum für Schwerionenforschung, Darmstadt, Germany}
\affiliation{$^7$III. Physikalisches Institut B, RWTH Aachen University, Aachen, Germany}
\affiliation{$^8$Institut für Kernphysik, Forschungszentrum Jülich, Jülich, Germany}

\begin{abstract}
We present first-principles numerical calculations of the depolarization rate of spin-polarized deuterium and tritium nuclei in realistic tokamak plasmas, driven by resonant interactions with plasma waves. Backed up by first-of-a-kind linear and nonlinear simulations, we find that alpha particle-driven Alfv\'enic modes cause only negligible depolarization, which is contrary to expectations in prior literature. Other Alfv\'enic instabilities can in principle degrade polarization, but only under conditions unlikely to be realized on transport timescales. By combining full-orbit particle tracing with a dedicated depolarization solver, we demonstrate that wave-driven depolarization is surprisingly weak in SPARC and ITER-scale devices. These results provide strong evidence that spin-polarized fuel can maintain its polarization long enough to boost fusion reactivity, opening a viable path toward substantially enhanced performance in magnetic confinement fusion power plants.
\end{abstract}

\maketitle

\section{Introduction}

For fusion power plants (FPPs) to become economically viable, they must sustain a sufficiently high volume-averaged fusion power density within the constraints of stability and engineering limits \cite{abdou1999exploring,Wade2021}. Conventional strategies to raise power density rely on increasing plasma density and temperature through improved confinement, external heating, and optimized magnetic configurations \cite{Lawson1957,Keilhacker1999,Creely2020,Wurzel2022,Austin2019,parisi2025a}.

An alternative and complementary strategy is to exploit the spin of the fuel nuclei. Polarizing the deuterium and tritium nuclear spins enhances the deuterium-tritium (D–T) fusion cross section \cite{Kulsrud1982,Cowley1984,Kulsrud1986,sandorfi2017polarized,Heidbrink2024}, thereby boosting fusion power density without increasing plasma pressure \cite{More1983,Mitarai1992,Temporal2012,ciullo2016nuclear,Smith2018IAEA,Heidbrink2024,Parisi_2025b}. This pathway is attractive because it bypasses stability and transport limits set by plasmas pressure, heating and particle sources \cite{Troyon1984,Greenwald1988,Horton1999,Jenko2001,Baker_2001,Ryter_2005,Camenen_2005,hender2007mhd,Mantica_2009,Hillesheim_2013,Stober_2000,Garbet_2004b,Holland_2013,Citrin_2014,Kim2024,parisi2025introduction}, while simultaneously increasing tritium burnup and improving tritium self-sufficiency \cite{Abdou2021,Meschini2023,Parisi_2024d}. Spin polarization also modifies the angular distribution of fusion products \cite{Kulsrud1982,Kulsrud1986,Baylor2023,Garcia2023,Garcia2025}, which could benefit tritium breeding, blanket transmutation, material longevity, and power conversion \cite{WhanBae_2025,Rutkowski2025,parisi2025revisiting}.

The promise of spin-polarized fuel (SPF) in FPPs is tempered by two key challenges: (i) ensuring that polarization survives long enough in the plasma to matter \cite{Kulsrud1986,Garcia2023,Baylor2023,Heidbrink2024}, and (ii) developing SPF sources with sufficient throughput for reactor fueling \cite{spiliotis2021ultrahigh,Baylor2023,reichwein2024plasma,Kannis2025}. This work addresses the first challenge of polarization lifetime. In particular, we focus on depolarization by plasma waves—a mechanism whose severity remains uncertain but represents one of the largest barriers to SPF deployment.

Experimental data on spin survival in fusion-relevant conditions are limited. The only direct measurement to date, using accelerated $^3$He ions, indicates that significant polarization can persist even at MeV energies \cite{zheng2024}. This, together with extensive theoretical studies \cite{Kulsrud1982,Coppi1983,More1983,Cowley1984,Kulsrud1986,goel1988spin,kulsrud2010polarization,paetz2016spin,gatto2016depolarization} and forthcoming experiments on the DIII-D tokamak \cite{buttery2019diii,Baylor2023,Garcia2023,Heidbrink2024,Garcia2025}, motivates computational exploration of depolarization in magnetically confined plasmas (MCF). While numerical studies exist for inertial confinement fusion \cite{hu2020spin,hu2023numerical}, no such work has yet been performed for MCF.

\begin{figure}[tb!]
    \centering
    \includegraphics[width=0.7\textwidth]{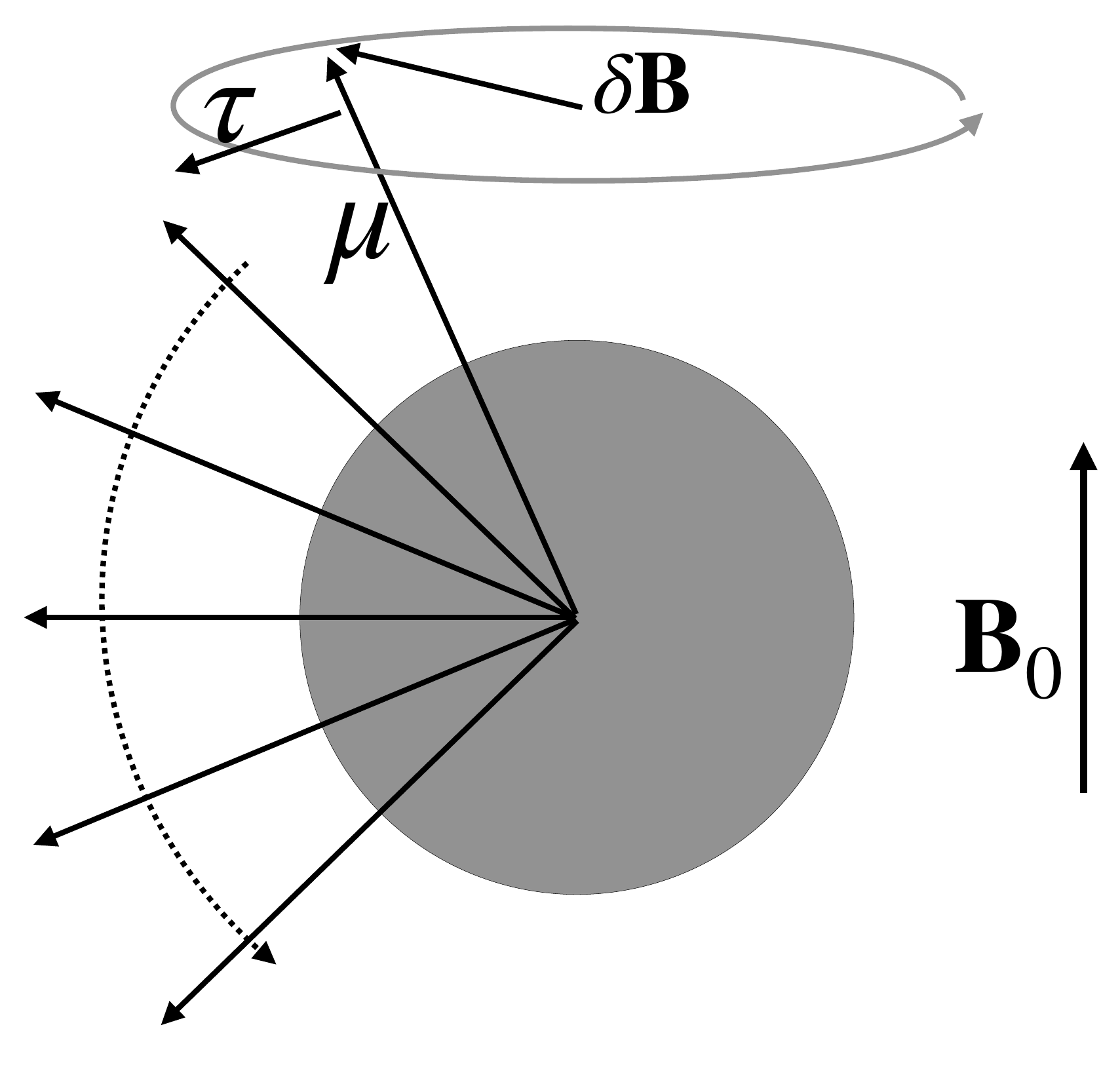}
    \caption{Schematic for how a circularly polarized wave with field $\delta \mathbf{B}$ exerts torque $\tau = \mathbf{\mu} \times \delta \mathbf{B}$ on a magnetic moment $\mathbf{\mu}$. If the wave and magnetic moment precession frequencies are close, there will be a torque for sufficiently long to change the component of $\mathbf{m}$ along the background field $\mathbf{B}_0$. Dashed arrow indicates the change of $\mathbf{\mu}$ over time.}
    \label{fig:spin_depolarization}
\end{figure}

\begin{figure*}[t!]
    \centering
    \includegraphics[width=0.95\textwidth]{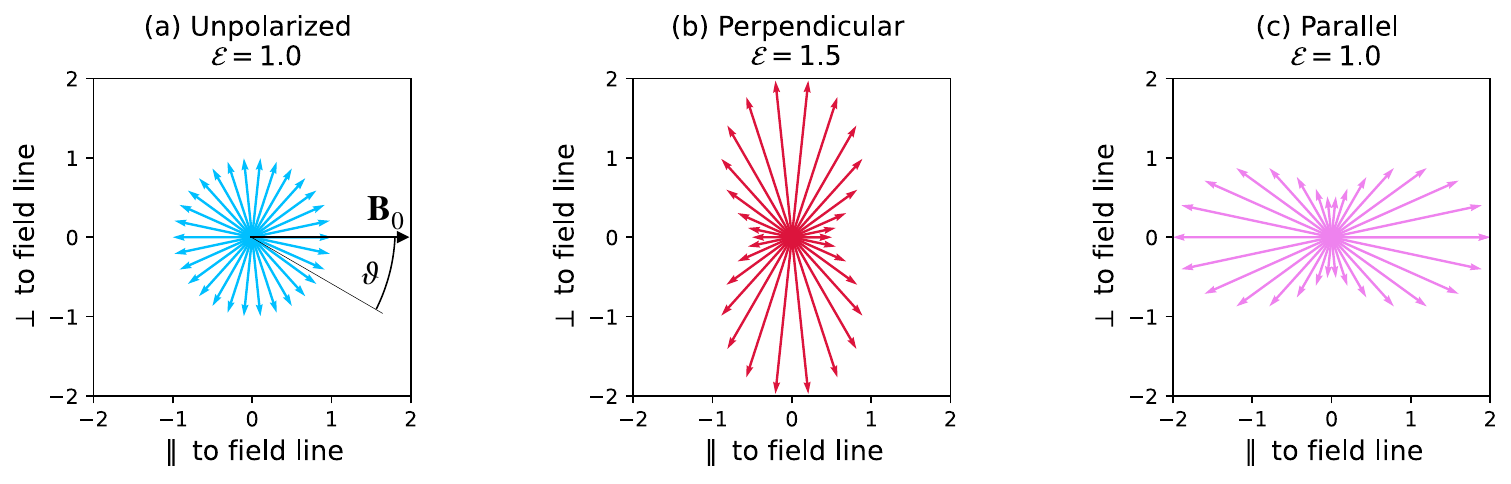}
    \caption{Relative density of neutron emission $F (\vartheta$) (see \Cref{eq:Fvartheta}) parallel and perpendicular to the magnetic field $\mathbf{B}_0$ for three spin-polarization schemes. In (a) we indicate the pitch angle $\vartheta$.}
    \label{fig:SPF_directions}
\end{figure*}

Depolarization in MCF plasmas may arise from many mechanisms including Coulomb collisions, wall interactions, or resonant wave–spin coupling. The first two mechanisms are generally benign under power plant conditions \cite{Kulsrud1986,eckstein1976reflection,greenside1984depolarization,wu2021wall,causey2002hydrogen,majeski2006enhanced,zakharov2007low,douai2015wall,maan2024estimates}. The most concerning pathway is resonance between plasma waves and the nuclear magnetic moment \cite{Heidbrink2024}. In a magnetic field of strength $B$, nuclei precess at the Larmor frequency
\begin{equation}
    \omega_L = \gamma_n B ,
\end{equation}
where $\gamma_n$ is the nuclear gyromagnetic ratio. Plasma waves with frequency $\omega \approx \omega_L$ can resonantly drive spin flips \cite{Coppi1983,Lodder1983,Cowley1984,Coppi1986,Kulsrud1986,Baylor2023,Heidbrink2024,Garcia2025}. Alfvén waves, ion-cyclotron waves, and ion-Bernstein waves occupy this frequency range and may represent a significant threat to SPF survival \cite{davidson1975electromagnetic,dendy1994excitation,gorelenkov1995alfven,fasoli1997alfven,brambilla1999numerical,heidbrink2008basic,cramer2011physics,garcia2011fast,dendy2015ion,chen2016physics,prokopyszyn2025confinement}.

In this paper we investigate cyclotron-frequency waves arising from several mechanisms including plasma normal modes, magnetoacoustic cyclotron instability (MCI), and compressional Alfv\'en eigenmodes (CAEs) \cite{Kolesnichenko1998,Fulop2000,Smith2003}. Cyclotron-frequency waves due to ion cyclotron emission (ICE), arising from MCI \cite{Belikov1976, Dendy1992PhysPlasB, Dendy1993, Dendy1994a, Kolesnichenko1998, Cook_2022}  are ubiquitous in magnetized plasmas and have been found in nearly all large magnetically confined fusion devices: tokamaks including TFR\cite{TFR_ICE_1978}, JET\cite{cottrell1988superthermal,cottrell1993ion,mcclements1999ion,cottrell2000identification,jacquet2011parasitic,mcclements2018observations}, TFTR\cite{cauffman1995alfvenic, dendy1995ion, cauffman1995ion_revsciinst}, ASDEX-Upgrade\cite{dinca2014ion, ochoukov2018core, ochoukov2018observations}, TUMAN-3M\cite{askinazi2018ion,askinazi2018spectrum}, DIII-D\cite{thome2019central,degrandchamp2022mode, crocker2022novel}, EAST\cite{liu2019ion, liu2020ion}, HL-2A\cite{tong2022development, liu2023identification}, JT-60U\cite{kimura1998alfven, ichimura2008observation, ichimura2008study, sato2010observation, sumida2017comparison, sumida2019characteristics}, KSTAR\cite{kim2018distinct, Thatipamula2016dynamic}, NSTX-U\cite{fredrickson2019emission,fredrickson2021chirping}; stellarators including LHD\cite{saito2009measurement, saito2013measurement}, W7-AS\cite{shalashov2003nbi}, W7-X\cite{moseev2021development}; and the cylindrical device LAPD \cite{Samant2025physics}.

ICE occurs when an inverted distribution of fast ions resonates with the fast Alfv\'en wave at frequencies at or near successive ion cyclotron harmonics of the energetic species. This instability is known to be excited by so-called shells\cite{dendy1993magnetoacoustic} of newly born alpha-particles as well as distributions comprising isolated regions of velocity space with well-defined parallel and perpendicular velocity, also known as ring-beams\cite{chapman2020comparing}. These velocity space distributions typically occur in the edge where banana orbits take marginally trapped particles on large radial excursions to the outboard edge (see Fig. 8 of Ref. \cite{dendy1995ion}), where they radiate so-called edge-ICE. Core-ICE can also be observed emanating from the core\cite{ochoukov2018core,ochoukov2018observations, ochoukov2019interpretation,liu2023identification}. ICE has also been reported from ohmic plasmas\cite{askinazi2018ion} providing evidence that some ICE signals may not be driven by fast ions. We show how polarized fuel gives rise to birth distributions of alpha-particles with more anisotropy than in the default unpolarized, and we go on to quantify any increase or decrease in growth rates of the MCI.

Here, we investigate wave-induced depolarization in the infinite homogeneous plasma limit to establish a baseline physical understanding and present calculations for realistic tokamak geometries. \Cref{sec:SPF} introduces SPF concepts and \Cref{sec:bloch_derivation} gives a simple picture of wave depolarization. \Cref{sec:depolarization_dynamics} describes the equations used to study depolarization. \Cref{sec:normal_modes} analyzes the plasma normal modes most relevant to resonant depolarization in the infinite homogeneous plasma limit, which partly informs the parameter space search for linearly unstable modes due to fusion product alpha-particles born from spin-polarized fuel ions, which are discussed in \Cref{sec:fast_particle}. Fully kinetic nonlinear studies in the infinite homogeneous plasma limit by particle-in-cell simulations in \Cref{sec:PIC} build upon the previous section and provide an understanding of the characteristics of the fast-particle excited waves. Next, ensembles of particles are tracked in zero dimensions and in three-dimensional a SPARC-like machine in \Cref{sec:particle_tracking}, providing some estimates of the potential of waves to depolarize fuel ions in an FPP. We conclude in \Cref{sec:discussion} with implications for the viability of SPF in future FPPs.

\section{Spin-Polarized Fuel} \label{sec:SPF}

In this section, we describe SPF focusing on D-T fuel. 

For a general D-T spin-polarization scheme, the differential cross section D-T cross section depends on the pitch angle $\vartheta$ relative to the magnetic field $\mathbf{B}$ \cite{Kulsrud1986},
\begin{equation}
\frac{d \sigma}{d \Omega} = \frac{\sigma_0}{4\pi} F(\vartheta),
\end{equation}
where $\sigma_0$ is the nominal unpolarized D-T cross section,
\begin{equation}
\begin{aligned}
F(\vartheta) \equiv 1 - \frac{a b}{2} + \frac{1}{2} \left( 3  a b \sin^2 \vartheta + \frac{c}{2} \left( 1 - 3 \cos^2 \vartheta \right)  \right),
\end{aligned}
\label{eq:Fvartheta}
\end{equation}
and the polarizations are 
\begin{equation}
\begin{aligned}
& a = d_1 - d_{-1} \in [-1,1], \\
& b = t_{1/2} - t_{-1/2} \in [-1,1], \\
& c = d_1 + d_{-1} - 2d_0 = 1 - 3 d_0 \in [-2,1],
\end{aligned}
\label{eq:abc_definitions}
\end{equation}
where $a$ and $c$ are the deuterium vector and tensor polarizations, $b$ is the tritium vector polarization, $d_m (m=-1,0,1)$ and $t_m (m = -1/2,1/2)$ are the probabilities for deuterium and tritium to occupy a spin state $m$ ($\sum_m d_m = \sum_m t_m = 1$), and the solid angle $\Omega$ is in Sterradians. 
The total cross section $\sigma$ is modified by a factor $\mathcal{E}$,
\begin{equation}
\sigma = \frac{\sigma_0}{4 \pi} \int_{\varphi = 0}^{2\pi} d \varphi \int_{\vartheta= 0}^{\pi} F(\vartheta) \sin \vartheta d \vartheta = \mathcal{E} \sigma_0,
\end{equation}
where $\varphi$ is the azimuthal angle. Note that $c$ affects $F(\vartheta)$ but never enters $\sigma$ explicitly. The polarization cross-section multiplier $\mathcal{E}$ is 
\begin{equation}
\mathcal{E} \equiv 1 + a b /2 ,
\label{eq:AJ2}
\end{equation}
and satisfies $\mathcal{E} \in [0.5,1.5]$ \cite{Kulsrud1986}. Unpolarized fuel has $\mathcal{E} = 1$. A 50\% cross-section enhancement ($\mathcal{E} = 1.5$) is obtained by choosing $a b = 1$ \cite{Kulsrud1986}, corresponding to aligning the deuterium and tritium nuclear spins.

In \Cref{fig:SPF_directions} we plot $F(\vartheta)$, indicating the relative density of neutron emission parallel and perpendicular to the magnetic field, for three spin-polarization schemes: (a) unpolarized, (b) perpendicular ($a = b = c = 1$), and (c) parallel ($a = b = 0, c = -2$). While the unpolarized scheme has isotropic neutron emission, the perpendicular scheme biases emitted neutrons perpendicular to the magnetic field and the parallel scheme biases parallel to the magnetic field. In this work, we will consider all three of these polarization schemes.

While the D-T fusion cross section is predicted to increase by as much as 50\%, the corresponding power increase in MCF devices has been reported to be even higher. Recent work found that with $\mathcal{E} = 1.5$, the total fusion power increased by 80-90\% \cite{Smith2018IAEA,Heidbrink2024} due to increased alpha heating. Similarly, in ICF the input energy requirement at constant gain is predicted to fall by nearly a factor of two \cite{pan1987spin}.

\section{Wave Depolarization Mechanism}
\label{sec:bloch_derivation}

In this section we describe how waves can depolarize nuclei. Nuclei with magnetic moments $\boldsymbol{\mu}$ and spin $\mathbf{I}$ in a magnetic field $\mathbf{B}(t)$ evolve according to
\begin{equation}
\begin{aligned}
      \frac{d\mathbf{M}}{dt} = & \gamma_n \mathbf{M}\times \mathbf{B}(t) - \frac{M_x\,\hat{\mathbf{x}} + M_y\,\hat{\mathbf{y}}}{T_2} \\ & - \frac{(M_z - M_z^{\mathrm{eq}})\,\hat{\mathbf{z}}}{T_1},
  \label{eq:bloch}
\end{aligned}
\end{equation}
where $\mathbf{M} \equiv \langle \mu_n \mathbf{I} \rangle$ is the ensemble magnetization, $M_z^{\mathrm{eq}}$ is the equilibrium value of the longitudinal magnetization, and $T_1$, $T_2$ are the longitudinal and transverse relaxation times. The polarization is preserved, since all spins precess together. We decompose the total magnetic field into a time-independent equilibrium component $\mathbf{B}_0$ and time-dependent fluctuating component $\delta \mathbf{B}_\perp(t)$,
\begin{equation}    
    \mathbf{B} = \mathbf{B}_0 + \delta \mathbf{B}.
\end{equation}
 The fluctuating component $\delta \mathbf{B}$ describes a plasma wave whose electromagnetic fields include a small transverse, circularly polarized magnetic component,
\begin{equation}
  \delta \mathbf{B}_\perp(t) = B_1 \bigl(\hat{\mathbf{x}}\cos\omega t + \hat{\mathbf{y}}\sin\omega t\bigr).
\end{equation}
In the laboratory frame, this field rotates at frequency $\omega$. In the rotating frame of the spin, if $\omega$ is sufficiently close to $\omega_L$, the wave’s field appears nearly static and exerts a steady torque on the nuclear spin precesses at the Larmor frequency $\omega_L = \gamma_n B_0$. Physically, the wave is continuously `nudging' the spin away from its original axis, just as in nuclear magnetic resonance. 

Each torque is tiny, because the spin’s Zeeman energy $\mu_n B_0$ is only $\sim 10^{-7}$~eV in a Tesla-scale field, but resonance allows the nudges to accumulate coherently. Over time, the spin 'tips` away from its aligned orientation, and repeated events across the ensemble randomize the polarization.

Applying the rotating-wave approximation \cite{allen2012optical} to \eqref{eq:bloch} in a frame rotating at $\omega$ shows that the spin precesses about an effective field
\begin{equation}
  \mathbf{B}_\mathrm{eff} = \frac{B_1}{2}\,\hat{\mathbf{x}} + \frac{\Delta}{\gamma_n}\,\hat{\mathbf{z}}, 
  \qquad \Delta \equiv \omega_L - \omega ,
\end{equation}
with the Rabi frequency $\Omega_R = \gamma_n B_1/2$. On resonance ($\Delta=0$), the spin is tipped at a rate $\Omega_R$, so that after a characteristic time $\tau_R \sim \pi/\Omega_R$ the spin can be completely flipped. Including dephasing, the wave-induced depolarization rate per nucleus is
\begin{equation}
  \Gamma_{\mathrm{wave}} \;\approx\; \Omega_R^2 T_2 \;=\; \frac{\gamma_n^2 B_1^2}{4}\,T_2 .
  \label{eq:Gamma-wave-coherent}
\end{equation}
Thus even a very small transverse wave field $B_1$ can cause significant depolarization if $T_2$ is long.

Plasma waves are not usually coherent but form a spectrum. For a stochastic transverse field $B_+(t) = \delta B_x(t)+i\delta B_y(t)$ with spectral density $S_{B_+}(\omega)$, the spin-flip rate generalizes to
\begin{equation}
  \Gamma_{\mathrm{wave}} \;=\; \frac{\gamma_n^2}{2}\, S_{B_+}(\omega_L).
  \label{eq:Gamma-spectral}
\end{equation}
Equation~\eqref{eq:Gamma-spectral} states that depolarization occurs in proportion to the wave spectral power at the Larmor frequency, with the correct handedness. Physically, only those modes whose magnetic field rotates in close phase with the spin can continuously torque it; all others average away.  
Depolarization can thus be viewed as a resonance problem: the nuclear spin is a tiny `antenna' tuned to $\omega_L$, and any plasma wave with matching polarization and frequency can drive it. The survival of fuel polarization in a fusion plasma requires that
\begin{equation}
  \Gamma_{\mathrm{wave}} \, \tau_E \ll 1,
\end{equation}
where $\tau_E$ is the confinement or fueling-replacement time. This condition highlights the challenge of maintaining nuclear spin polarization in the presence of waves: ensuring that the plasma spectrum has little power near $\omega_L$ and its Doppler-shifted harmonics.

\section{Depolarization Dynamics} \label{sec:depolarization_dynamics}

In this section, we introduce the equations for studying depolarization of deuterium and tritium nuclei. This framework will form the foundation for inquiry in subsequent sections in configurations relevant to fusion machines.

The time-dependent Schr\:odinger equation describes the evolution of the deuterium or tritium wavefunction $\vec \psi$, which describes the evolution of nuclear spin,
\begin{equation}
\frac{d {\vec \psi}}{d t} = - \frac{i}{\hbar} {\mathcal H} {\vec \psi},\label{eq:schro}    
\end{equation}
where ${\mathcal H}$ is the Zeeman Hamiltonian. Electric dipoles of deuterons and tritons have not been measured and are not included. The Zeeman Hamiltonian is
\begin{equation}
{\mathcal H} = -g \mu_N {\vec B}_k S_{ijk}\label{eq:hamiltonian}
\end{equation}
where $g = 0.8574$ ($5.9579$) is the g-factor of the deuteron (triton), $\mu_N = 5.0509 \times 10^{-27} J/T$ is the nuclear magneton, and ${\vec B}_k$ is the magnetic field vector and $S_{ijk}$ is a Spin tensor. Normalizing away factors of $\hbar$, the components of $S_{ijk}$ for deuterium, and spin-1 particles in general, are
\begin{eqnarray}
S_{ijx} &=& \begin{bmatrix} 0 & 1 & 0 \\ 1 & 0 & 1 \\ 0 & 1 & 0 \end{bmatrix}\frac{1}{\sqrt{2}},\\
S_{ijy} &=& \begin{bmatrix} 0 & -i & 0 \\ i & 0 & -i \\ 0 & i & 0 \end{bmatrix}\frac{1}{\sqrt{2}},\\
S_{ijz} &=& \begin{bmatrix} 1 & 0 & 0 \\ 0 & 0 & 0 \\ 0 & 0 & -1 \end{bmatrix}.
\end{eqnarray}
and the equivalent tensors for tritons (and spin 1/2 particles in general) are
\begin{eqnarray}
S_{ijx} &=& \frac{1}{2}\begin{bmatrix} 0 & 1 \\ 1 & 0 \end{bmatrix},\\
S_{ijy} &=& \frac{1}{2}\begin{bmatrix} 0 & -i \\ i & 0 \end{bmatrix},\\
S_{ijz} &=& \frac{1}{2}\begin{bmatrix} 1 & 0 \\ 0 & -1 \end{bmatrix}.
\end{eqnarray}
The reference coordinate system for spin alignment is the z-axis, the direction of the static magnetic field $\vec B_0$. 

The non-commutative nature of the operators\cite{pauli1927quantenmechanik} make it difficult to find analytical solutions to the spin Hamiltonian dynamics. They may only be found for limiting cases, such as for low field strengths, exact resonances or far from resonance. Magnus expansions\cite{magnus1954on} alleviate this problem but generate large and unwieldy expressions that retain these limitations to some extent. In this work, we choose to use a high-order ODE solver which bypasses this problem and allows for the spin dynamics to be calculated simultaneously with charged particle motion in a magnetic field. For the remainder of this section we focus only on the spin dynamics of stationary particles.

\begin{figure}[bt]
    \centering
    \includegraphics[width=0.95\textwidth]{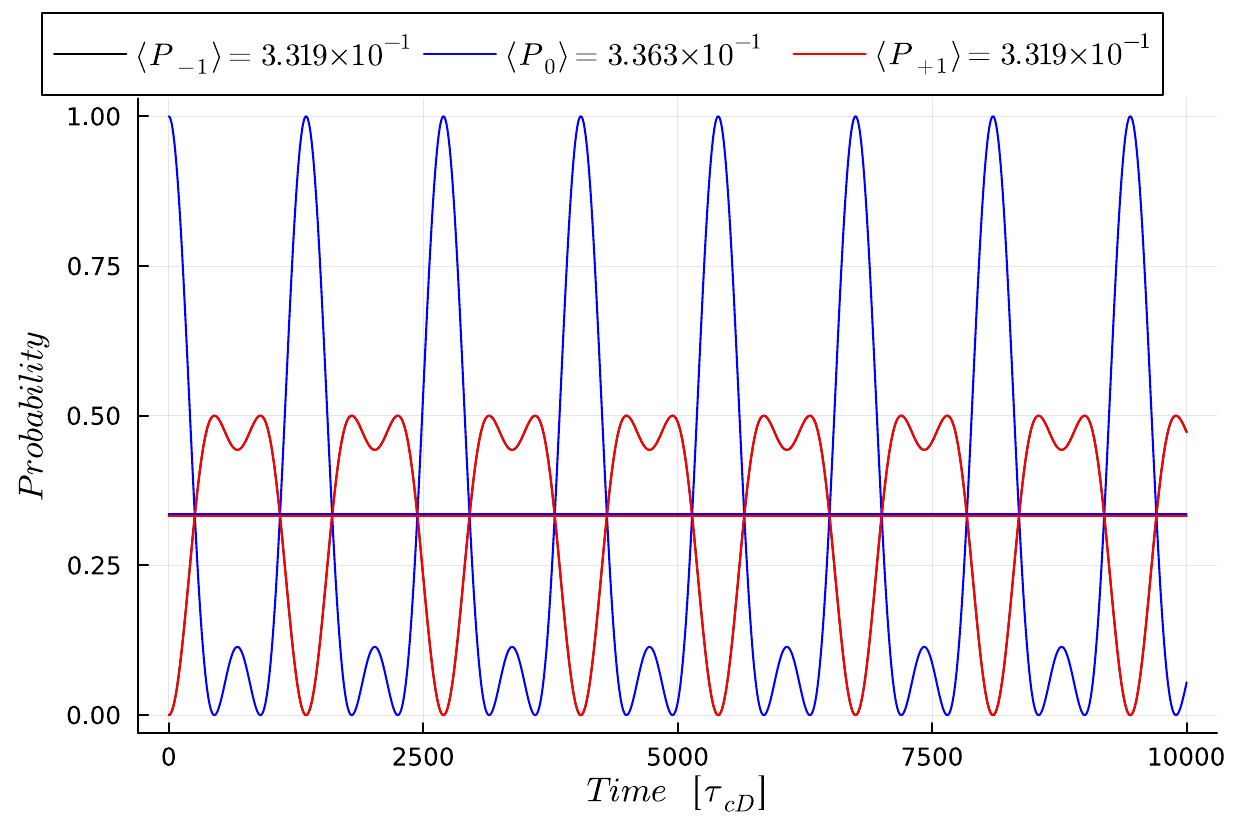}
    \caption{Time evolution of the probability density $\mathbf{\rho}$ (see \Cref{eq:rho_eq}) of deuteron spin states over the course of $10,000$ cyclotron periods in the presence of a left handed magnetic wave field with a frequency equal to the deuteron precession frequency and an amplitude $1/1000$ that of the bulk field. The curved traces represent the time evolution of the state probability and the straight lines show the temporal mean probability, values of which are listed in the legend.}\label{fig:temporalprobabilityresonance}
\end{figure}

The equation of motion of the spin Hamiltonian is integrated numerically using the `Vern9'\cite{verner_98_coefficients} ordinary differential equation time-stepping algorithm with absolute and relative tolerances of $10^{-8}$. The state vector of deuterium is a normalized 3-element complex vector $\ket{u}$,
\begin{equation}
    \ket{u} = c_{-1} \ket{-1} + c_{0} \ket{0} + c_{1} \ket{1},
\end{equation}
where $|c_{-1}|^2 + |c_{0}|^2 + |c_{1}|^2 = 1$, and $\ket{-1}$, $\ket{0}$, and $\ket{1}$ are the basis vectors representing the instantaneous spin states $m_I \in [-1, 0, 1]$.
In this work, we measure the spin states over time by calculating the instantaneous expectation values of $\ket{u}$ in the $i^{th}$ direction, $\bra{u}S_i \ket{u}$, and (ii) the vector $\mathbf{\rho}$ of occupational probability of each spin state,
\begin{equation}
    \mathbf{\rho} = \mathrm{diag} \left( \ket{u}\bra{u} \right).
    \label{eq:rho_eq}
\end{equation}
Figure \ref{fig:temporalprobabilityresonance} shows the time trace of $\rho$ of these three states for a left handed magnetic perturbation with an amplitude $10^{-3}$ times less than the bulk static field. Electric field perturbations play no part and are not included. The wave frequency is exactly the deuteron Larmor precession frequency. Note that the initial condition is $[0, 1, 0]'$ meaning the spin is exactly in the $m_I=0$ state. Over the course of 100s of cyclotron periods, the deuteron depolarizes nearly completely to an average state of $\sim 1/3$ probability each, although the depolarization model features Rabi oscillations\cite{rabi1937space} whereby the spin states pass through their initial conditions.

\begin{figure}[bt!]
    \centering
    \includegraphics[width=0.95\textwidth]{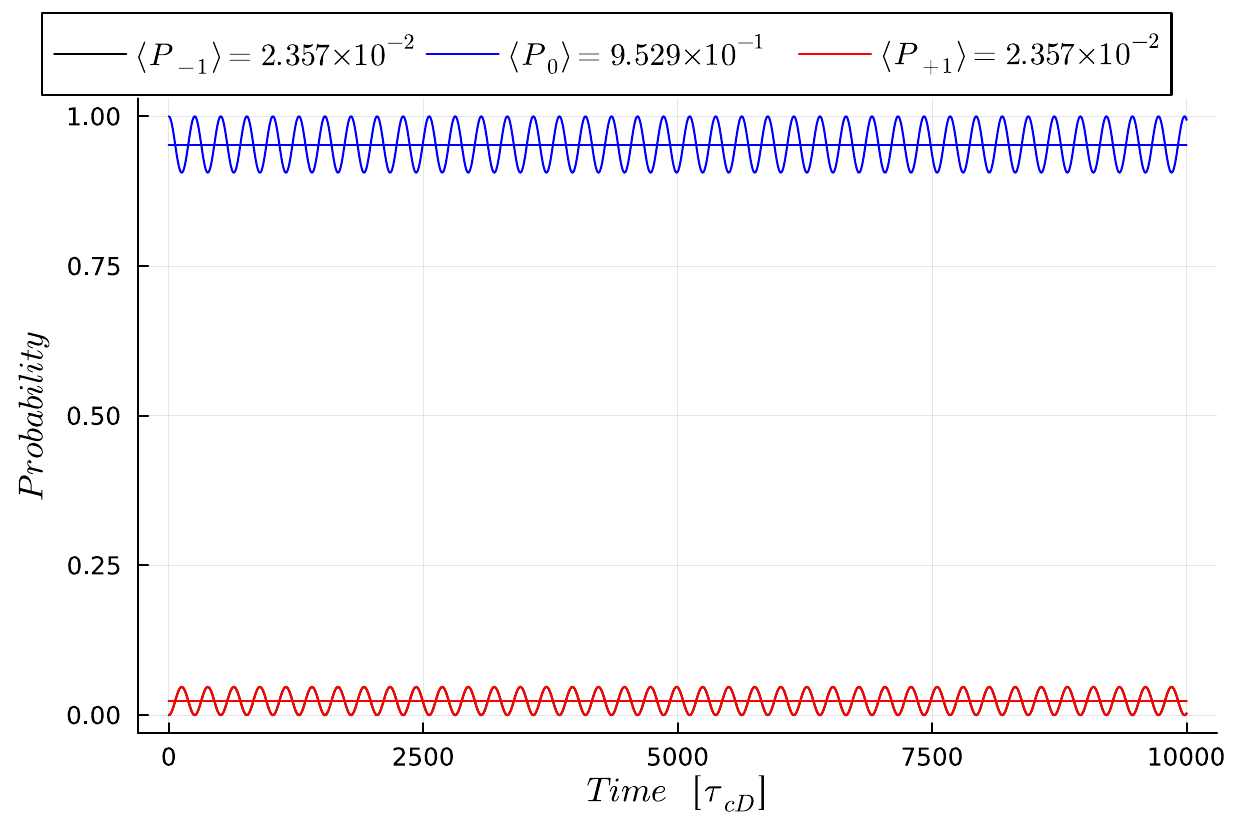}
    \caption{As in Fig. \ref{fig:temporalprobabilityresonance} except the wave field has a frequency $0.995$ that of the the deuteron precession frequency.}
    \label{fig:temporalprobabilityoffresonance}
\end{figure}

When the frequency of the applied wave field moves further from resonance, the amplitude of the Rabi oscillations decreases but the frequency increases, as shown in Figure \ref{fig:temporalprobabilityoffresonance}.

There are two crucial observations to be made at this point: first, only relative magnetic perturbations perpendicular to the bulk magnetic field contribute to depolarization; second, only the left-handed (right-handed) component of the perpendicular magnetic perturbation can depolarize particles with positive (negative) Larmor precession frequencies.

\begin{figure}[bt!]
    \centering
    \begin{subfigure}[t]{0.96\textwidth}
    \centering
    \includegraphics[width=1.0\textwidth]{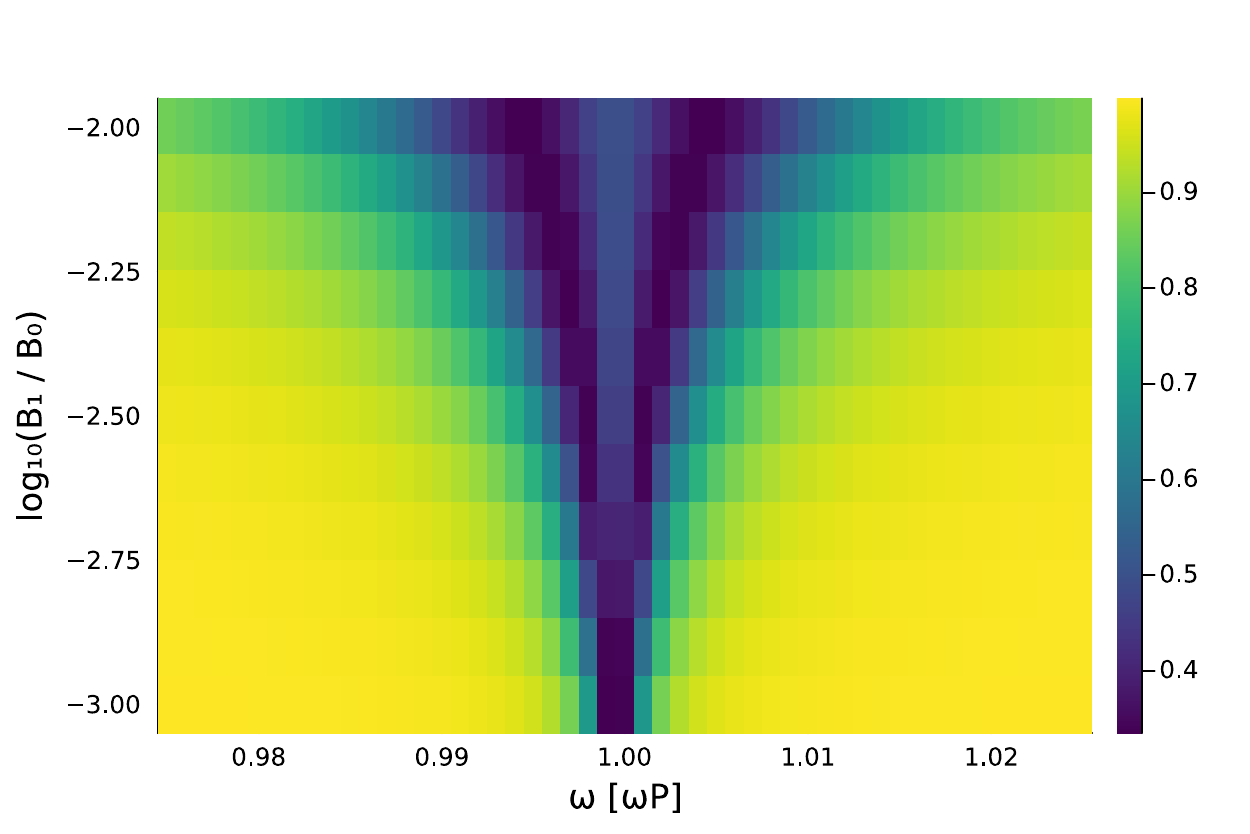}
    \caption{The probability of finding a deuteron in spin state $m_I=0$ after $10,000$ cyclotron periods.}
    \label{fig:uniformscan0}
    \end{subfigure}
    \centering
    \begin{subfigure}[t]{0.96\textwidth}
    \centering
    \includegraphics[width=1.0\textwidth]{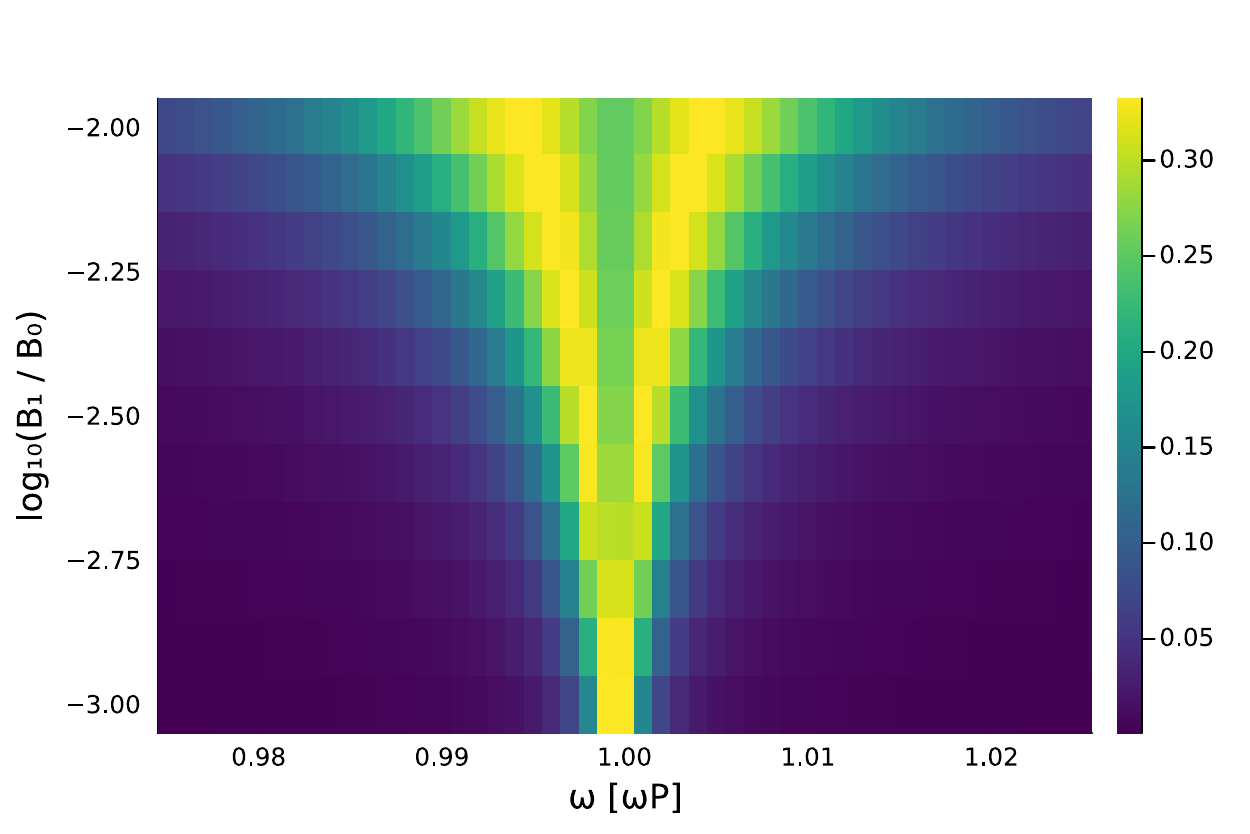}
    \caption{The steady state probability of finding a deuteron in spin state $m_I=1$ after $10,000$ cyclotron periods. The result for $m_I=-1$ is identical.}
    \label{fig:uniformscan1}
    \end{subfigure}
    \caption{Shading indicates the probability of finding a deuteron in a particular spin state after $10,000$ cyclotron periods in the presence of a left-handed magnetic field wave with varying relative amplitudes and frequencies. The y-axis indicates the scan in relative amplitudes of wave field to bulk magnetic field on a $log_{10}$ scale. The x-axis shows the frequencies in units of the deuteron Larmor precession frequencies. The deuteron is initially in $m_I=0$. A probability of 1/3 for $m_I$ indicates complete depolarization.}
    \label{fig:FBE_general}
\end{figure}

To investigate this further, we perform scans of depolarization for left-handed magnetic field waves of varying amplitudes and frequencies. Figs. \ref{fig:uniformscan0} and \ref{fig:uniformscan1} show how the temporal mean probability changes of a deuteron that is initially in the $m_I=0$ state. 

Figs. \ref{fig:uniformscan0} and \ref{fig:uniformscan1} show that the resonance width of the depolarization process is narrower at lower amplitudes and that depolarization is negligible for relative wave amplitudes of $10^{-4}$. Both the amount of depolarization and the width of the resonance increases with wave amplitude. Relative wave amplitudes of $10^{-2}$ are unlikely, but allow us to see the general trend of the structure of depolarization. Except when ion cyclotron resonance heating is deployed, FPPs are unlikely to encounter waves of amplitudes that pose a danger to polarized fuel. Regardless, we investigate this possibility in later sections.

The inhomogeneous bulk magnetic field of an FPP changes both the cyclotron frequency and Larmor precession frequency. As such, the spatial locations where a wave may resonantly depolarize a fuel ion are are localized, which for tokamaks are vertical stripes due to the $1/R$ dependence of the toroidal field. Particles must dwell within these regions in order to undergo coherent interactions leading to depolarization, hence only a subset of orbits are at risk.

In summary, we have introduced the formalism used for studying nuclear depolarization and demonstrated the effects of wave depolarization for several cases. We now proceed to study one of the mechanisms that leads to potentially depolarizing waves.

\section{Normal modes} \label{sec:normal_modes}

\begin{figure*}[t!]
    \centering
    \includegraphics[width=1\linewidth]{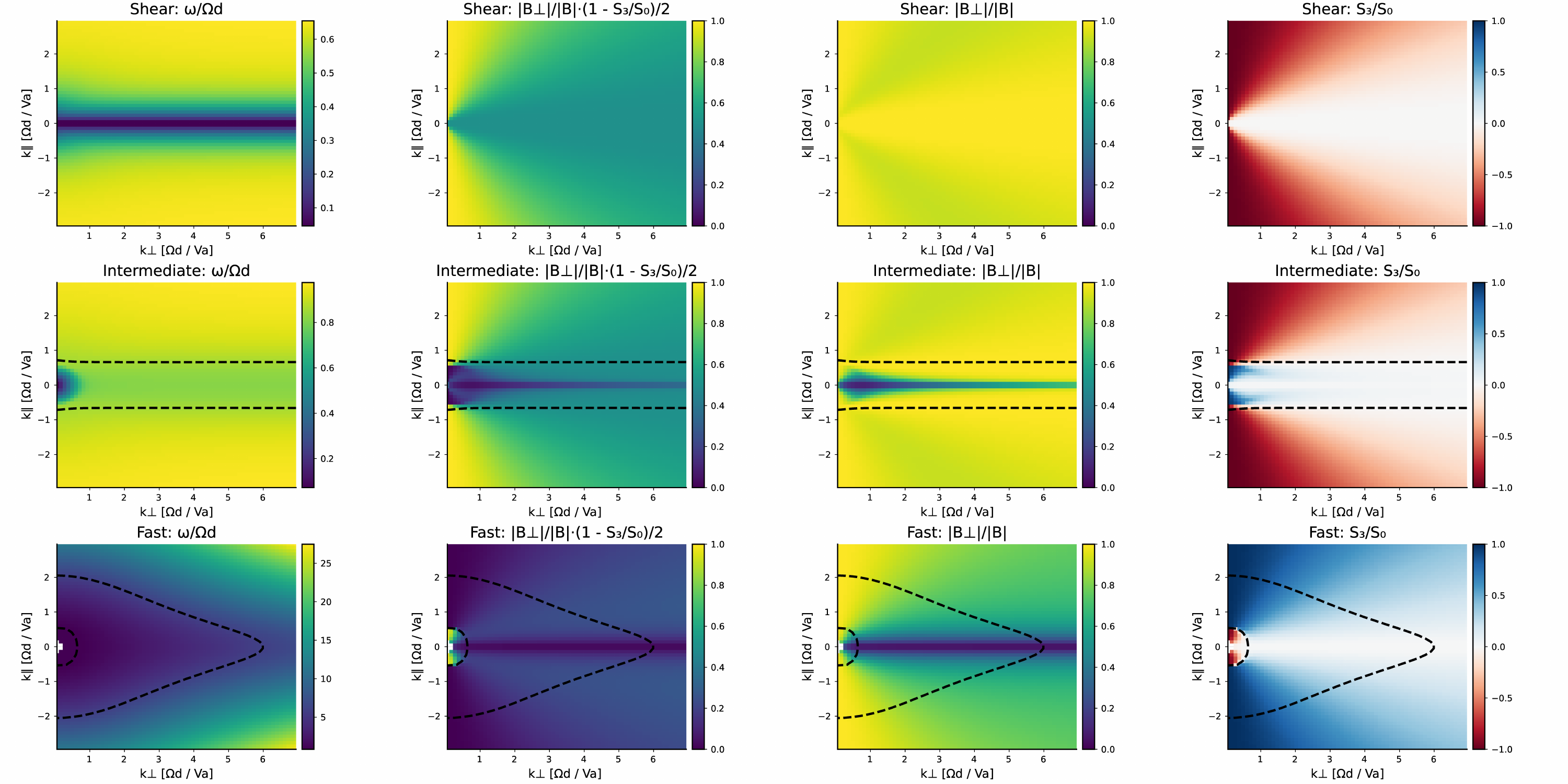}
    \caption{Normal modes of a 50-50 DT mix plasma in a parallel and perpendicular wavenumber plane. The three rows indicate three different normal modes of the system in order of lowest to highest frequency form top to bottom: the shear Alfv\'en wave; intermediate or hybrid mode; fast Alfv\'en wave. Shading  of the four columns show different information about the normal modes, from left to right: frequency in units of the deuteron cyclotron frequency; the amount of perpendicular magnetic perturbation that is left handed, see Eq. \ref{eq:perplefthand}; the relative fraction of perpendicular magnetic perturbation; the handedness of the wave, $+1$ ($-1$) is right (left) handed. The dashed black traces indicate frequency iso-contours at the deuteron and triton Larmor precession frequency, which are $\sim0.86$ and $\sim5.9$ respectively. The semi-circular black dashed linear in the lower right panel separates the red from blue region by chance - it does not do so for other DT ratios.}
    \label{fig:normalmodes}
\end{figure*}

In this section, we study normal plasma modes that could depolarize D-T fuel. Normal modes of the plasma at frequencies near the deuterium and tritium precession frequencies are of particular interest for D-T mixes where we need to be aware of which may be detrimental to polarization. Focusing first on exploring them in the cold plasma limit, we find the dispersion surfaces of the waves that are supported by a 50-50 D-T mix plasma. From these surfaces we obtain the wave frequency and handedness with respect to the bulk magnetic field.

The quantity of interest regarding depolarization of an arbitrary wave is the amount of perpendicular perturbed amplitude that is left-hand polarized. Stokes' parameters measure the polarization of an electromagnetic perturbation where $S_3 \in [-1, 1]$, when normalized by the amplitude $S_0$, and indicates the handedness where normalized $S_3 = + 1$ is right-handed and normalized $S_3=-1$ is left-handed. The relative amplitude of the perpendicular wave magnetic field is $|B_{1\bot}|/|B_1|$. Hence the product of these two quantities,

\begin{equation}
\frac{|B_{1\bot}|}{|B_1|} \left(1 - \frac{S_3}{S_0}\right)\frac{1}{2} \in [0, 1],
\label{eq:perplefthand}
\end{equation}

\noindent measures the ability of a wave to depolarize a deuterons and tritons. Figure \ref{fig:normalmodes} introduces these quantities for the 3 normal modes near the deuteron and triton Larmor precession frequencies in a 50-50 DT mix plasma.

It is clear from reading the upper row of Fig. \Cref{fig:normalmodes} that modes on the shear Alfv\'en branch do not reach the deuteron precession frequency and can be considered safe in terms of their capability to depolarize deuteron fuel. The intermediate mode, middle row, with small $|k_\perp|/|k|$ may depolarize deuterons while waves on the fast Alfv\'en branch, lower row, in the same wavenumber limit may depolarize both deuterons and tritons. Whether these modes are excited or not is another matter, which we turn to in the next section.


\section{Fast-Particle Instabilities} \label{sec:fast_particle}

In this section we investigate the degree to which alpha-particles born from different polarized fuel schemes are unstable to the MCI and hence capable of causing ICE, which matters in this context for two very important reasons. The first reason is that it has the potential to emit waves at both the deuteron and triton Larmor precession frequency, endangering the polarization of both species. The second reason is that polarization schemes alter the birth distribution of alpha-particles in pitch, resulting in greater anisotropy and therefore giving the distribution more free energy. In this section we quantify, for the first time, the difference in the MCI growth rates of alpha particles born from isotropically distributed fuel spins and that of the distribution function arising from $a=b=0$ and $c =-2$ and $a=b=c=1$.

\begin{figure}[tb]
    \centering
    \includegraphics[width=0.9\linewidth]{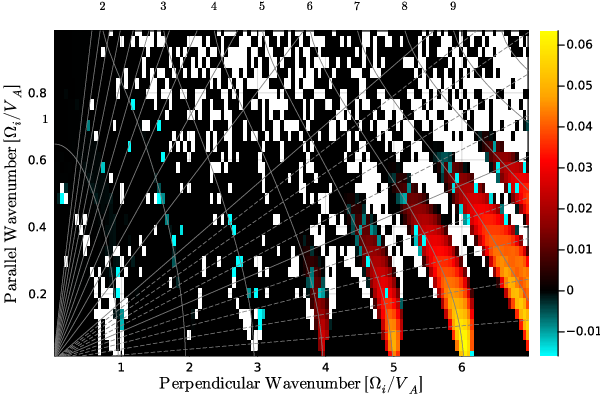}
    \caption{Shading  indicates the growth rates of the unstable modes on the fast Alfv\'en dispersion surface on a two dimensional 64-by-64 parallel and perpendicular wavenumber grid. The unstable modes are driven by the anisotropic alpha-particle velocity space distribution, Eq. \ref{eq:alphaparticlemodelf}, born from spin-polarized DT reactions with spin parameters $a= b=c=0$. Here $v_{th} = v_0/1000$. White pixels in the figure indicate where the algorithm was not able to find a solution.}
    \label{fig:growthrates000}
\end{figure}

The pitch $p$ of alpha-particles born from polarized D-T reactions can be drawn from a distribution function
\begin{equation}
\begin{aligned}
g_{a,b,c}(p) = & \frac{1}{2} \left(1 - \frac{a b}{2}\right) + \\
&\frac{1}{2} \left(3 a b (1 - p^2) + \frac{c}{2} (1 - 3 p^2)\right),
\label{eq:alphafpitch}
\end{aligned}
\end{equation}
where $a$, $b$, and $c$ are defined in \Cref{eq:abc_definitions}. Unpolarized fuel, $a = b = c = 0$, results in isotropically distributed alpha particles on the surface of a sphere in three velocity space dimensions i.e. a shell distribution. The particles are all born with 3.6 MeV.

In the following numerical analysis we modify this function by a Maxwellian thermal speed
\begin{equation}
\begin{aligned}
& f_{a,b,c}(v_z, v_\bot) = \\  & g_{a,b,c}\left(\frac{v_z}{\sqrt{v_z^2 + v_\bot^2}}\right) \exp\left(-\frac{(\sqrt{v_z^2 + v_\bot^2} - v_0)^2}{v_{th}^2}\right).
\end{aligned}
\label{eq:alphaparticlemodelf}
\end{equation}
We find that the analysis is insensitive to changes to $v_{th}$ if it is $\frac{v_0}{100}$ or less, where $v_0$ is the speed corresponding to 3.6 MeV. Larger values of $v_{th}$ are not physical and we would require a slowing-down distribution of the source of particles drawn from Eq. \ref{eq:alphafpitch}.

The distribution function $f_{a,b,c}$ is passed into the linearized Maxwell-Vlasov set of equations\cite{Stix2ndEd1992, Brambilla_1998, Swanson_2003}, along with Maxwellian electrons, deuterons and tritons, in order to calculate its complex frequency solutions for sets of given $(k_z, k_\bot)$ pairs: see Eqs. 5-9 of Ref. \cite{Cook_2022}. The linearized system describes the self-consistent growth or attenuation of an electromagnetic wave oscillating according to the collective motion of the plasma species, where charge and current perturbations are self-consistently coupled to the electromagnetic wave that co-exists with them. Maxwell's equations \ref{eq:monopoles}-\ref{eq:ampere} determine the evolution of the electric field $\vec E$ and magnetic field $\vec B$, the Vlasov equation \ref{eq:vlasov} and moments of the 6-dimensional distribution function $f_s$ of species $s$ with charge $q_s$ and mass $m_s$ that provide the charge density $\rho$, Eq. \ref{eq:charge}, and current density $\vec J$, Eq. \ref{eq:current}, as sources. 

\begin{eqnarray}
\nabla\cdot \vec B &=& 0\label{eq:monopoles}\\
\nabla\cdot \vec E &=& \frac{\rho}{\epsilon_0}\label{eq:gauss}\\
\frac{\partial {\vec B}}{\partial t} &=& -\nabla \times {\vec E}\label{eq:faraday}\\
\mu_0 \epsilon_0 \frac{\partial {\vec E}}{\partial t} &=& \nabla \times {\vec B} - \mu_0 \vec{J}\label{eq:ampere}\\
\frac{\partial f_s}{\partial t} &=& - {\vec v}\cdot \nabla_x f_s - \frac{q_s}{m_s}(\vec E + \vec v \times \vec B)\cdot\nabla_v f\label{eq:vlasov}\\
\epsilon_0 \rho &=& \sum_s{\int dv q_s n_s f_s}\label{eq:charge}\\
\vec J &=& \sum_s{\int dv \vec v q_s n_s f_s}\label{eq:current}
\label{eq:maxwellvlasov}
\end{eqnarray}

We use LinearMaxwellVlasov.jl\cite{Cook_2022, Slade-Harajda_2025} to find these solutions after linearization. It makes efficient use of the Lerche-Newberger rules\cite{Lerche_1966,Newberger_1982,Swanson_2003} to convert the standard truncated sum over cyclotron harmonics for integrals in $v_\parallel$ and $v_\perp$ into a single integral and hence computationally tractable in a two-dimensional $(k_z, k_\bot)$ plane.

The default depolarized fuel, $a=b=c=0$, growth rates of the magnetoacoustic cyclotron instability are shown first in the following three figures \ref{fig:growthrates000}, \ref{fig:growthrates00-2}, and \ref{fig:growthrates111}. The color scale shows the growth rate in units of alpha-particle cyclotron frequency, and white indicates where the algorithm failed to find a solution. The x-axis (y-axis) is the parallel (perpendicular) wavenumber in units of $\Omega_{D} / V_A$. Curved traces indicate contours of the real part of the complex frequency solution from which we obtain growth rates: the integers on the left and upper sides of the plot indicate the value of the real frequency in units of $\Omega_D$. Straight dashed lines indicate angles of constant propagation in degrees off perpendicular, and solid lines indicate the same in steps of 5 degrees. These plots are symmetric about $k_\parallel = 0$ and only the upper half-plane in parallel wavenumber is plotted.


The case where $a=b=0$ and $c=-2$ shows the same mode structure with growth rates approximately $2/3$ that of the unpolarized case:

\begin{figure}
    \centering
    \includegraphics[width=0.9\linewidth]{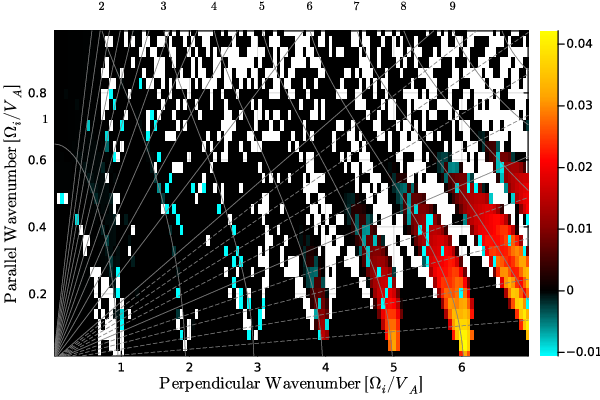}
    \caption{As in \ref{fig:growthrates000} except $a=b=0$ and $c=-2$.}
    \label{fig:growthrates00-2}
\end{figure}


Finally, for $a=b=c=1$ the growth rates are approximate 20\% larger than the unpolarized case.

\begin{figure}
    \centering
    \includegraphics[width=0.9\linewidth]{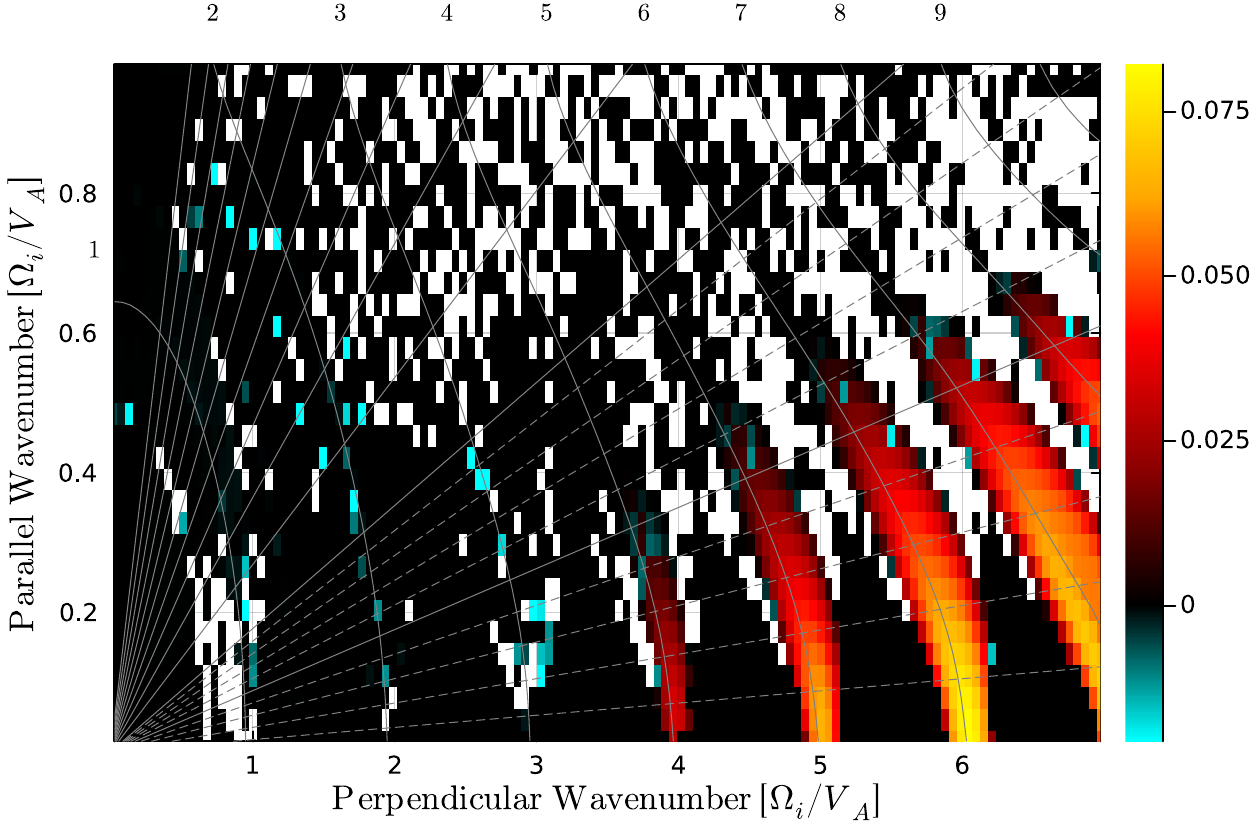}
    \caption{As in \ref{fig:growthrates000} except $a=b=c=1$.}
    \label{fig:growthrates111}
\end{figure}


\begin{figure*}[tb]
    \centering
    \includegraphics[width=0.99\linewidth]{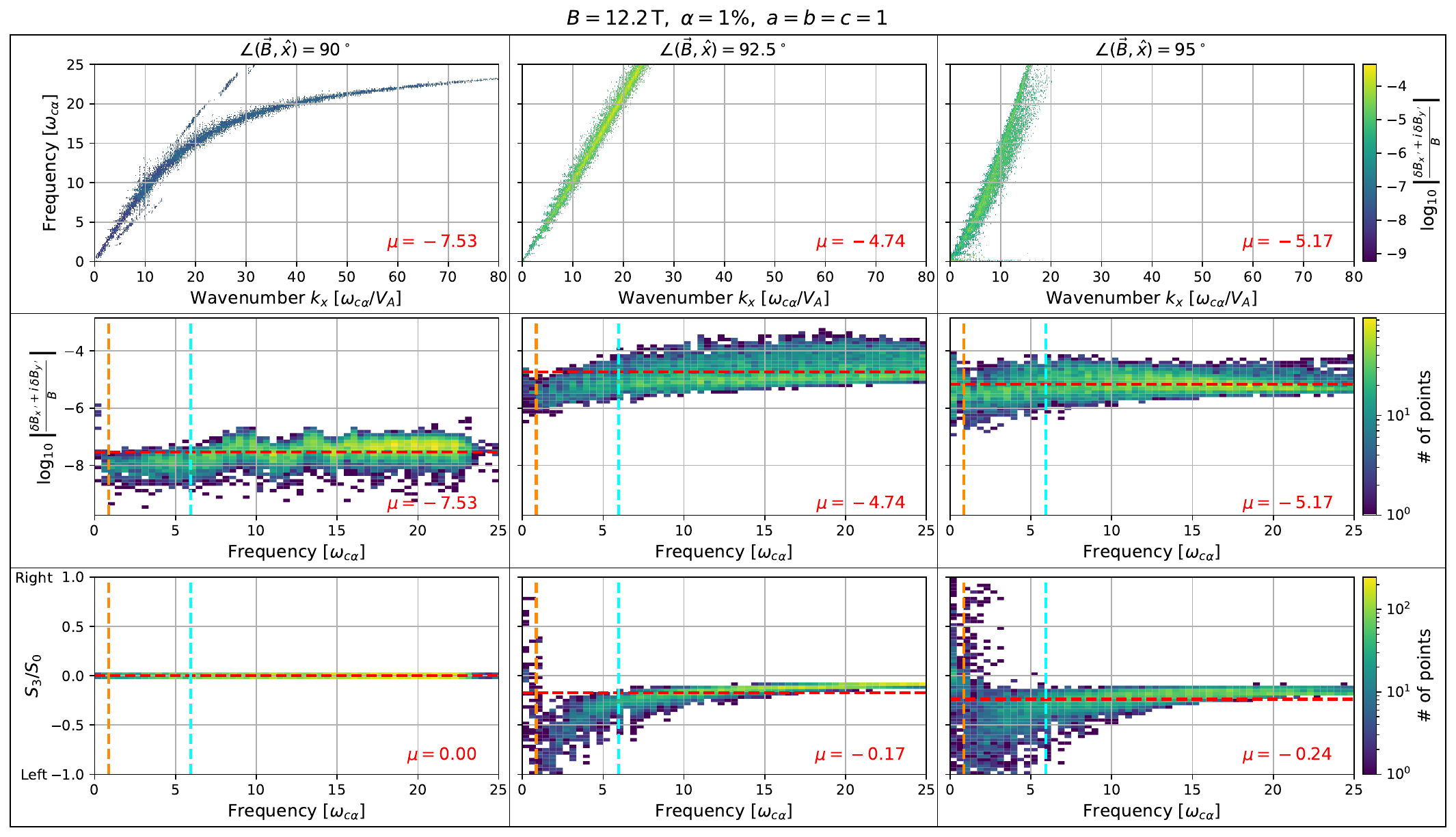}
    \caption{EPOCH simulation results for SPARC at three magnetic field orientations relative to the simulation domain ($\angle(\vec{B},\hat{x}) = 90^\circ,\,92.5^\circ,\,95^\circ$, shown column-wise), with constant alpha fraction ($\alpha = 1\%$) and fuel polarization ($a=b=c=1$). Color in the top row shows the $\log_{10}$ of the absolute value of the circular polarization component of the magnetic field waves, normalized to the bulk magnetic field, as a function of frequency against wavenumber. The middle row shows two-dimensional histograms of the data plotted in the upper panels as a function of the magnitude of the circular polarization component of the magnetic wave fields against frequency. The lower panels are similar to the middle panel except the ordinate is the handedness Stokes parameter of the waves. The orange and cyan vertical dashed lines indicate the deuteron and triton precession frequencies, respectively. On all plots $\mu$ shows the mean value of the quantity plotted in each panel.}
    \label{fig:SPARC_angle}
\end{figure*}

We see that for this background density of $10^{20}$ m$^{-3}$ and magnetic field of $6$ T, the unstable spectrum is located on and around the 4-7 alpha-particle cyclotron harmonics and that propagation is $\pm 6^o$ around perpendicular. Note that there is no instability at the deuteron Larmor precession frequency below the first cyclotron harmonic but the $6^{th}$ harmonic, near the triton Larmor precession frequency, is unstable. However, Referring to the cold-plasma dispersion surface plot originating from the cold-plasma description, we see that these waves have very little left-handed perpendicular perturbations making them benign in terms of their ability to depolarize the fuel.

Given that ICE is not regularly observed from unpolarized fuel alpha-particles before they have slowed down, this suggests that ICE from fuel polarized alpha-particles will not present a massive threat given their growth rates are broadly similar.

Non-linear interactions between ICE waves at adjacent higher harmonics couple power to the lower harmonics, including the fundamental\cite{Chapman_2020,Samant2025physics}. This could be a mechanism where ICE power is cascaded down to the deuteron Larmor precession frequency where the left-handed perpendicular magnetic perturbation is no longer negligible.

\section{Particle-in-Cell Simulations} \label{sec:PIC}

In this section, we describe particle-in-cell (PIC)\cite{Birdsall_1985, Villasenor_1992} simulations used to model fusion born alpha-particle driven instabilities in fusion plasma conditions. PIC has been used to model the MCI associated with energetic ion distributions where, importantly, it takes the physics into the nonlinear phase allowing power emitted from linearly unstable high harmonics to couple to lower harmonics\cite{cook2013particle, toida2018simulation, ochoukov2019interpretation, chapman2019interpretation, reman2019NF,chapman2020comparing,samant2024predicting,Samant2025physics}: it is this coupling that offers an extra energy transfer channel from fast ions to waves at and near the Larmor precession frequencies of the fuel ions.

Introducing spin-polarized fuel in fusion reactors results in an anisotropic distribution of fusion products which can excite distinct magnetic perturbations in a plasma. Here we use PIC simulations in one spatial dimension and three velocity space dimensions that self-consistently evolve Maxwell's equations and particles under the Lorentz force using the EPOCH code \cite{Arber:2015hc}. We estimate the size of magnetic perturbations driven by the anisotropic alpha particle distribution in a spin-polarized 50-50 D-T plasma. The standard EPOCH source code is modified to incorporate the initial alpha particle distribution function corresponding to various values of $a$, $b$, and $c$, as described in equation \ref{eq:alphafpitch}. Each simulation is evolved for a duration of eight alpha cyclotron periods. The single spatial domain aligned along the $x$-axis whilst the background magnetic field $\mathbf{B}$ has a fixed magnitude and lies in the $x$-$z$ plane. The direction of $\mathbf{B}$ relative to the $x$-axis, fuel polarization ($a,b,c$), and alpha fraction ($\alpha$) are varied across simulations. The simulations are run for SPARC ($T = \SI{7}{\kilo\electronvolt}$, $n = 3 \times 10^{20}\,\mathrm{m}^{-3}$, $B = \SI{12.2}{\tesla}$) and ITER ($T = \SI{15}{\kilo\electronvolt}$, $n = 1 \times 10^{20}\,\mathrm{m}^{-3}$, $B = \SI{5.3}{\tesla}$), where $T$ and $n$ are the plasma temperature and density respectively.

Magnetic field fluctuations transverse to $\mathbf{B}$ can be expressed as the superposition of left-hand and right-hand circularly polarized components, defined relative to an axis parallel to $\mathbf{B}$. Depolarization takes place only when the wave rotates in the same direction as the ion is precessing. The left-hand component is of particular importance (in case of positive Larmor precession frequency) as it can interact resonantly with nuclear spin precession; a frequency match between the two provides a direct channel for spin depolarization. We obtain these magnetic field wave characteristics from EPOCH simulations as a function of frequency and wavenumber via Fourier transform. The wavenumber $k$ is defined by the axis of variation in EPOCH, i.e. the $x$-axis. Through the parameter scan in angle between the magnetic field and the $x-axis$, the plane perpendicular to the magnetic field changes and hence a rotation around the $y$-axis is deployed to obtain the wave information in a coordinate system $(x',y'=y,z')$ where the magnetic field points exactly along $z'$. This ensures consistent extraction of perpendicular magnetic fluctuations $\delta B_{x'}$ and $\delta B_{y'}$, from which the relative left-hand component $\left| (\delta B_{x'} + i\,\delta B_{y'}) / B \right|$, as well as the Stokes ratio $S_3/S_0 = -2\,\mathrm{Im}(\delta B_{x'} \,\delta B_{y'}^{\dagger}) \, / \, (|\delta B_{x'}|^2 + |\delta B_{y'}|^2)$, are subsequently calculated.

\begin{figure*}[t!]
    \centering
    \includegraphics[width=1\linewidth]{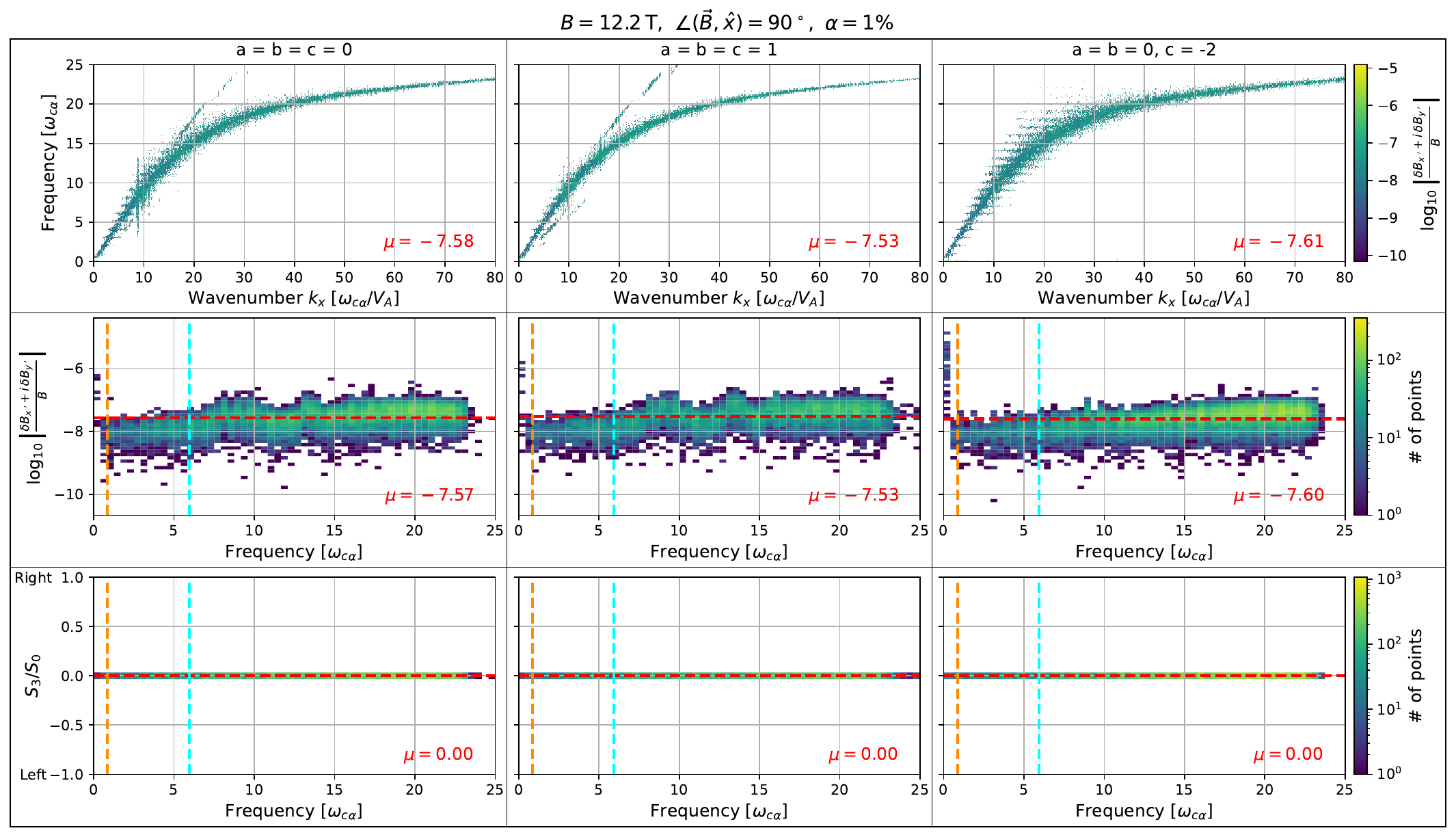}
    \caption{EPOCH simulation results for SPARC at three fuel polarization configurations ($a=b=c=0$, $a=b=c=1$, and $a=b=0,\,c=-2$, shown column-wise), with constant magnetic field orientation $\angle(\vec{B},\hat{x}) = 90^\circ$, and alpha fraction $\alpha = 1\%$.}
    \label{fig:SPARC_fuel}
\end{figure*}

\begin{figure*}[t!]
    \centering
    \includegraphics[width=1\linewidth]{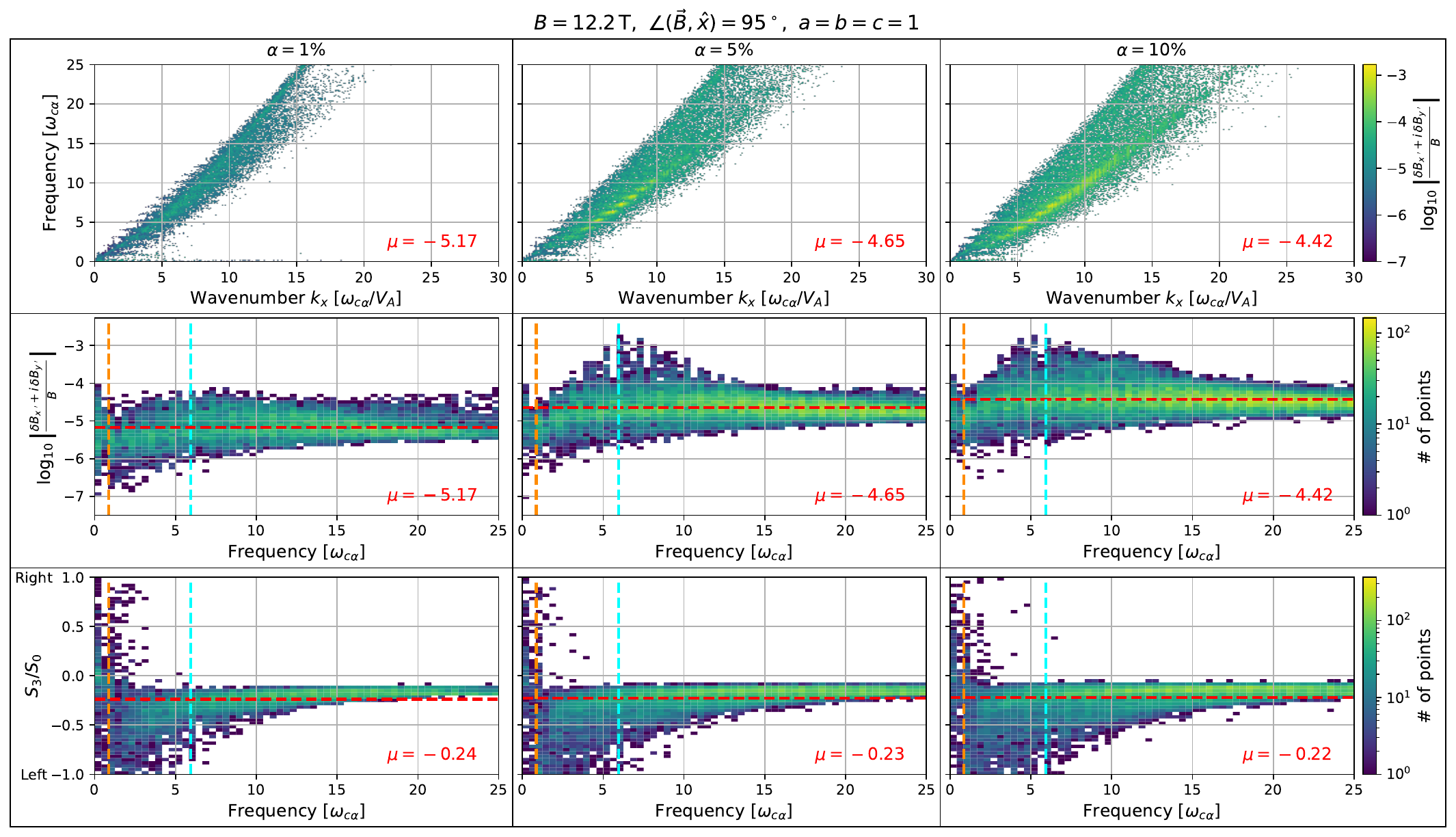}
    \caption{EPOCH simulation results for SPARC at three alpha fractions ($\alpha = 1\%,\,5\%,\,10\%$, shown column-wise), with constant magnetic field orientation $\angle(\vec{B},\hat{x}) = 95^\circ$, and fuel polarization $a=b=c=1$.}
    \label{fig:SPARC_alpha}
\end{figure*}

The simulation results are presented for the SPARC \cite{Creely2020} (Figures \ref{fig:SPARC_angle}, \ref{fig:SPARC_fuel}, \ref{fig:SPARC_alpha}) and ITER (Appendix \ref{sec:ITER}) tokamaks. The frequency and wavenumber axes are normalized to the alpha cyclotron frequency, $\omega_{c\alpha}$, and $\omega_{c\alpha}/V_A$, respectively, where $V_A$ is the Alfv\'en speed. For each subplot, the mean value ($\mu$) of the relevant quantity is indicated by a horizontal red dashed line and annotated with red text. Vertical dashed lines mark the precession frequencies of deuterium (orange) and tritium (cyan), which occur at approximately $0.86\,\omega_{c\alpha}$ and $5.94\,\omega_{c\alpha}$, respectively \cite{Kulsrud1986}.

Based on all the simulation results, it can be seen that the primary magnetic field fluctuations in the $(\omega,k)$ spectra correspond to the fast Alfv\'en branch, with superposed discrete ion-cyclotron resonances (ICR). This result indicates that Alfv\'en waves, rather than ion–cyclotron instabilities, constitute the dominant threat to sustaining spin polarization. In figure \ref{fig:SPARC_angle}, as the angle $\angle(\vec{B}, \hat{x})$ shifts from perpendicular to slightly oblique ($92.5^\circ$ and $95^\circ$), the emergence of a non-zero parallel wavenumber ($k_\parallel$) increases the mean magnetic field fluctuations amplitude by two to three orders of magnitude and drives a transition from linear to left-handed polarization. At oblique angles, the wave polarization transitions from left-handed ($S_3/S_0<0$) at low frequencies to nearly linear ($S_3/S_0 \approx 0$) at higher frequencies, in contrast to the purely linear polarization observed at $90^\circ$. 

The magnetic fluctuation amplitude of the combined Alfvén–ICR branch is largely insensitive to the choice of fuel polarization $(a,b,c)$, with both the amplitude spectra and the Stokes parameter ratio $S_3/S_0$ remain nearly unchanged across all polarization schemes, as illustrated in figure \ref{fig:SPARC_fuel}. Increasing the fast-alpha fraction from $\alpha=1\%\to 5\%\to 10\%$ raises the mean level from $\mu=-5.17\to-4.65\to-4.42$, while the mean polarization ($S_3/S_0$) remains weakly left-handed and nearly $\alpha$–independent as shown in figure \ref{fig:SPARC_alpha}. Similar results are also obtained for ITER (Appendix \ref{sec:ITER}).

\section{Particle Tracking in Realistic Geometry} \label{sec:particle_tracking}

In this section, we integrate our spin-depolarization code with realistic particle trajectories in a real machine.

One may integrate several trajectories originating from across the phase space and perform a Monte Carlo integral of the bulk expected depolarization state $\bar {<\psi_0>}$ over an evolution time $\tau$, where
\begin{widetext}
\begin{equation}
\centering
 \langle \overline{ \psi}_0 \rangle = \frac{\sum_{v_\parallel}\sum_{v_\bot}\sum_{V} {\bar \psi}_0(v_\parallel, v_\bot, V) v_\bot \Delta V\Delta v_\parallel \Delta v_\bot}{\sum_{v_\parallel}\sum_{v_\bot}\sum_{V} v_\bot \Delta V\Delta v_\parallel \Delta v_\bot}.
\end{equation}
\end{widetext}

\begin{figure}[h!]
\centering
\includegraphics[width=1\linewidth]{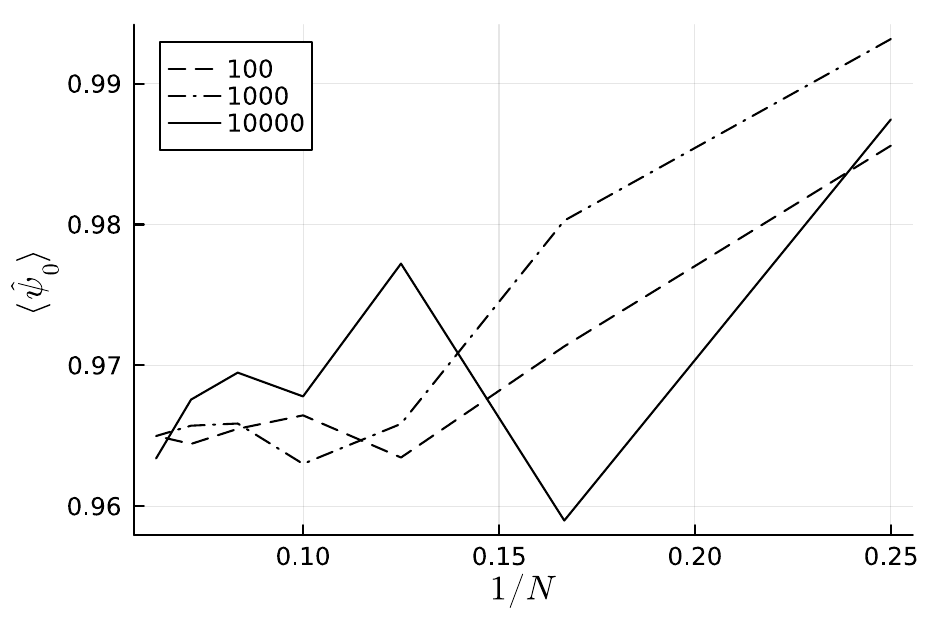}
\caption{The spatio-temporal expectation value of the $m_I=0$ state, initially fully occupied, evaluated from an ensemble of particles integrated over 10,000 cyclotron periods (as calculated on axis) for a SPARC-like equilibrium. Here $N$ is the number of locations in minor radius, parallel velocity and perpendicular speed at which particles are initialized in order to perform the Monte-Carlo integral of spatio-temporal expectation spin state.}
\label{fig:sparcdepolarizationzerok}
\end{figure}

Figure \ref{fig:sparcdepolarizationzerok} displays a convergence study for such a calculation on a SPARC-like machine, beginning with all particles in the $m_I=0$ state in the presence of a left-handed wave with zero wavenumber and amplitude $10^{-3}$ that of the field on axis. The wave frequency is that of the precession frequency on-axis. Deuterons are drawn uniformly from an axial temperature on-axis of $100$ eV, which falls off quadratically in $s = \sqrt{1-\psi/\psi_{ax}}$ towards the last closed flux surface (LCFS), where $\psi$ is the poloidal magnetic flux and $\psi_{ax}$ is that on axis. A resolution study has been performed in the linear scan size $N$, where $N$ is the number of equally spaced samples in the integration dimensions; flux surfaces, parallel velocity and perpendicular velocity. The particles are tracked for 10,000 on-axis cyclotron periods. Initially, the deuterons are all in the $m_I=0$ state and we see that the temporal mean state sees a reduction of depolarization by $\sim 3.5\%$. The other two states split the remaining probability equally between them.

One would need to run these with high values of $N$ for more accurate results and include collisions to allow particles to traverse phase space whilst going into and out of resonance. Despite this, one may optimistically draw a conclusion from these initial data that depolarization may be negatively correlated with higher temperatures; large Larmor radii serves to delocalize the ions with respect to spatially isolated regions of resonance. The next subsection accounts for non-zero wavenumber and multiple co-existing waves.

\subsection*{Bulk spin depolarization by eigenmodes in SPARC}
The bulk spin depolarization of an ensemble of initially polarized deuterium or tritium ions is evaluated in the SPARC axisymmetric equilibrium magnetic field—modeled using a Solov’ev equilibrium with Miller shaping (major radius $R_{0} = 1.85 \text{m}$, on-axis toroidal field $B_{0} = 12.2 \text{T}$, inverse aspect ratio $\epsilon = 0.31$, elongation $\kappa = 1.97$, and triangularity $\delta = 0.54$)—in the presence of superimposed eigenmode perturbations (see figures \ref{fig:D1}, \ref{fig:D0}, and \ref{fig:T1}). The use of compressional Alfv\'en eigenmode (CAE)-like structures\cite{Kolesnichenko1998,Fulop2000,Smith2003} is motivated by the active area of study on the link between higher frequency MCI fast Alfv\'en modes that give good agreement from infinite homogeneous plasma models and the geometry dependent cyclotron and sub-cyclotron eigenmodes of Hall-MHD\cite{Hellsten2003,Smith2009compressional, sharapov2014bi,bradshaw2020modelling}. The plasma volume is radially discretized into $N_{v}=10$ approximately equal-volume nested toroidal shells, $\{\Delta V_i\}_{i=1}^{N_v}$, representing the volume between adjacent magnetic flux surfaces fixed by the normalized minor-radius coordinate $s_{i} = \sqrt{1-\psi_{i}/\psi_{\text{ax}}}$, where $\psi_{i}$ is the poloidal flux on the $i$-th surface and $\psi_{\text{ax}}$ is its value at the magnetic axis, such that $s=0$ at the axis and $s\approx1$ at the LCFS. Ions are initialized on the outboard midplane of each flux surface $s_i$ at major radius $R_i$. At each of these spatial locations, the velocity space is sampled on a tensor grid of $N_E \times N_E$ points (where $N_E=15$) in parallel and perpendicular kinetic energy relative to the equilibrium magnetic field, $(E_{\parallel,j},E_{\perp,k})$. The grid points are quadratically spaced from 0 to $E_{\max}=12\,\text{keV}$ according to the relation $E_n = E_{\max}u_n^2$, where $\{u_n\}_{n=1}^{N_E}$ is drawn from a uniform distribution in $[0,1]$. This non-uniform sampling provides a higher density of grid points at low energies, which better represents the bulk of a thermal Maxwellian distribution while sparsely sampling the high-energy tail. For each discrete initial condition (position $R_i$ and energy $\{E_{\parallel,j}, E_{\perp,k}\}$), the evolution of a single spin polarized ion is simulated over collisionless timescales. The particle's trajectory is determined by the Lorentz force, while its spin state, $|\psi(t)\rangle$, evolves according to the time-dependent Schrödinger equation, yielding the probability of the ion being in each specific spin state $m_I$ as a function of time $P_{m_{I},ijk}(t)$. The final bulk probability is then determined by scaling this $P_{m_{I},ijk}(t)$, by a weighting factor, $W_{ijk}$, calculated as follows to represent its statistical significance within the overall particle distribution
\begin{equation}
W_{ijk} = (1 - s_i^2) \, e^{-\tfrac{E_{\parallel,j}}{T}} \, e^{-\tfrac{E_{\perp,k}}{T}} \, \sqrt{E_{\perp,k}} \, \Delta V_i.
\end{equation}
Here $(1-s_i^2)$ provides a parabolic on-axis-peaked density profile, the Maxwellian factors $\exp(-E/T)$ weight the energy grid at a common temperature $T=1\,\text{keV}$ (identical in $\parallel$ and $\perp$), a term from the velocity-space volume element, $v_{\perp,k} \propto \sqrt{E_{\perp,k}}$, and $\Delta V_i$ converts the surface sum into a volume integral. The final bulk probability for a specific spin state $m_I$ as a function of time, $\langle P_{m_I}(t) \rangle$, is given by the weighted average over the entire simulated particle distribution:
\begin{equation}
\langle P_{m_I}(t) \rangle = \frac{\sum_{i,j,k} P_{m_I,ijk}(t) \cdot W_{ijk}}{\sum_{i,j,k} W_{ijk}}  
\end{equation}

This summation is the numerical approximation of the continuous integral over the ions' phase-space distribution function.

\begin{figure}
    \centering
    \includegraphics[width=0.9\textwidth]{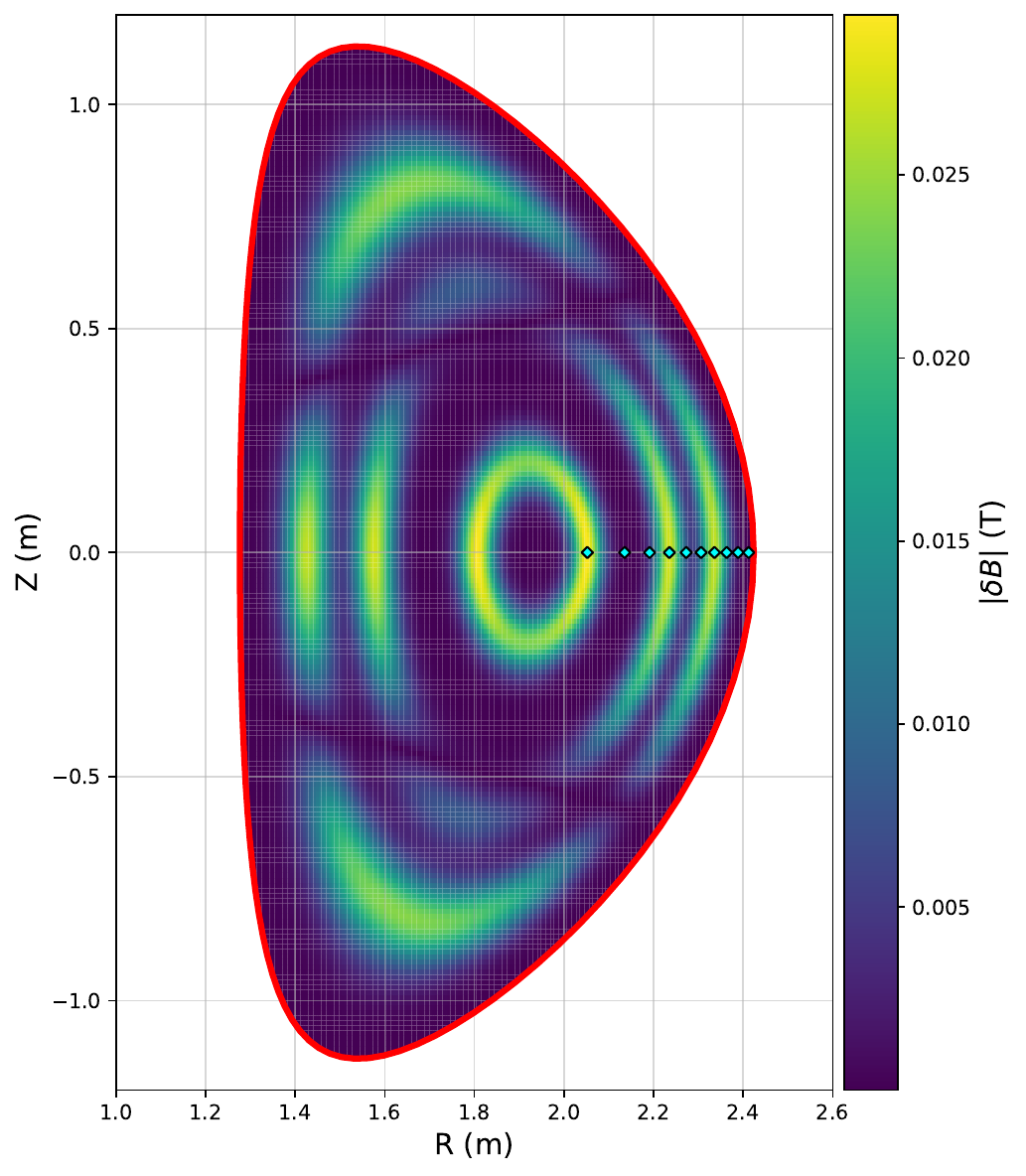}
    \caption{Magnitude of eigenmode perturbations $|\delta B|$ in the SPARC tokamak plasma at $t=0$, shown in the $(R,Z)$ cross-sectional plane. The red line indicate the plasma boundary, and the cyan colored diamonds mark the initial positions of ions on the outboard midplane.}
    \label{fig:eigenmodes}
\end{figure}

\begin{figure}
    \centering
    \includegraphics[width=0.9\textwidth]{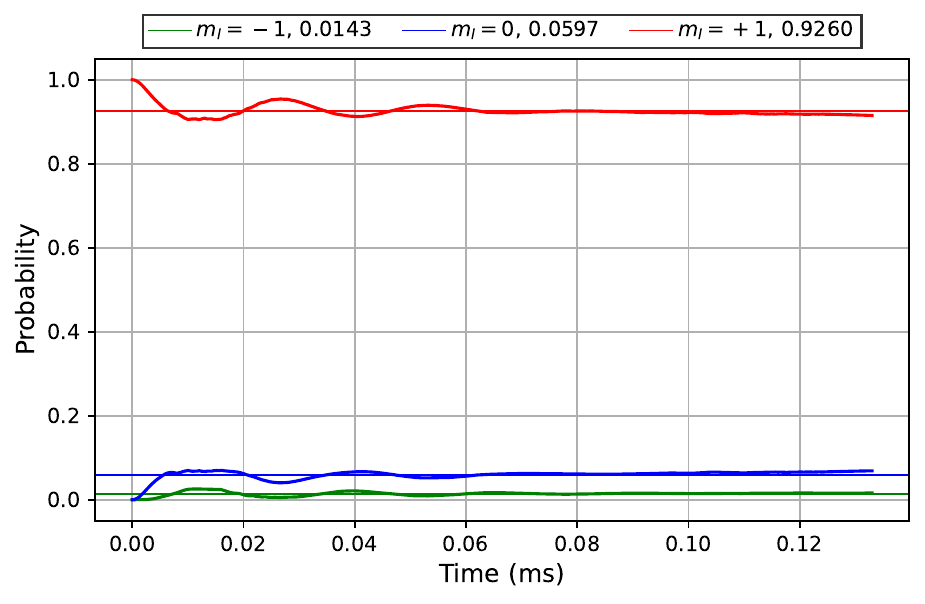}
    \caption{Time evolution of the bulk probability density of deuterons spin states in SPARC, initialized in the spin polarized state $m_I=1$, over collisionless timescales in the presence of eigenmode perturbations. The thick curved traces represent the time evolution of the spin state probability, while thin straight lines indicate their temporal mean values, listed in the legend.}
    \label{fig:D1}
\end{figure}

\begin{figure}
    \centering
    \includegraphics[width=0.9\textwidth]{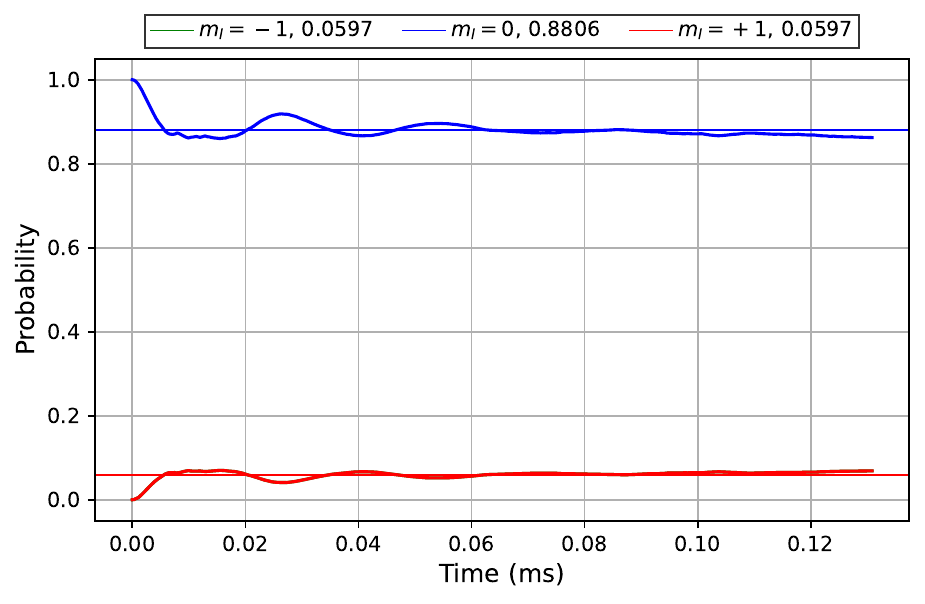}
    \caption{Time evolution of the bulk probability density of deuterons spin states in SPARC, initialized in the spin polarized state $m_I=0$, over collisionless timescales in the presence of eigenmode perturbations. The thick curved traces represent the time evolution of the spin state probability, while thin straight lines indicate their temporal mean values, listed in the legend.}
    \label{fig:D0}
\end{figure}

\begin{figure}
    \centering
    \includegraphics[width=0.9\textwidth]{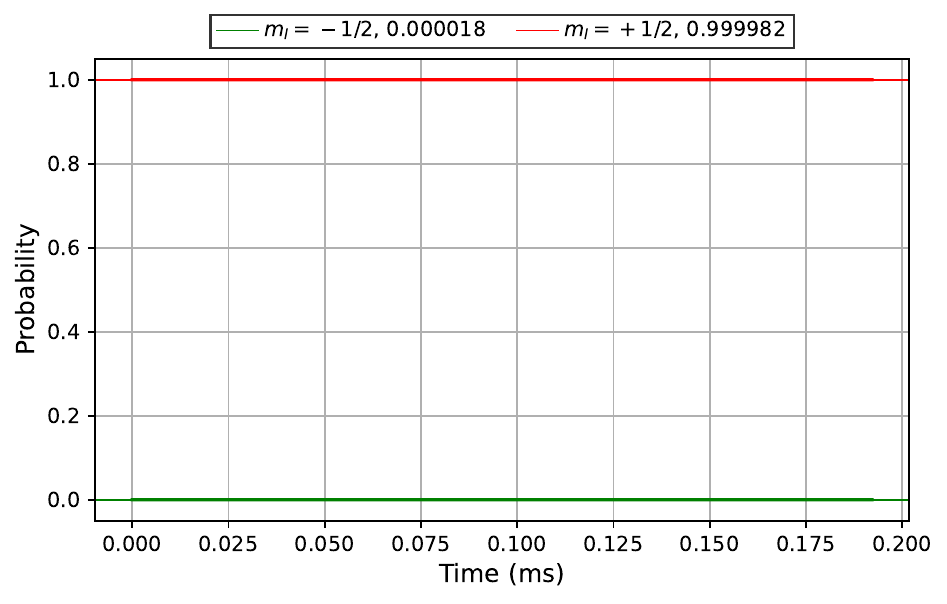}
    \caption{Time evolution of the bulk probability density of tritium spin states in SPARC, initialized in the spin polarized state $m_I=+1/2$, over collisionless timescales in the presence of eigenmode perturbations. The thick curved traces representing the time evolution of the spin state probability nearly overlap with their temporal mean values lines, indicating negligible temporal variation (mean values in legend).}
    \label{fig:T1}
\end{figure}

Eigenmode perturbations at a given position and time (see figure \ref{fig:eigenmodes}) are modeled as a superposition of $l=3$ eigenmodes, expressed as a divergence-free magnetic fluctuation of the form:
\begin{equation}
\delta\mathbf{B}(s,\theta,\phi,t) \;=\; 
  \Re\!\left[\sum_{l=1}^{3} \hat{\mathbf{B}}_{\perp}^{l}\, e^{\,i\Phi_{l}}\right]   
  \label{eq: modes_sum}
\end{equation}

where the phase of each mode is $\Phi_l = [\,k_s^{l} s \,+\, m_l\theta \,+\, n_l\phi \,-\, \omega_l t\,]$, with $k_s^{l}=0$, $m_l=6$, and $n_l=3$ denoting the radial, poloidal, and toroidal mode numbers, respectively. Each mode is localized around a selected flux surface $s_i$ ($i=1,4,7$) and its angular frequency $\omega_l$ is set to the local deuteron (or triton) precession frequency, which is evaluated using the equilibrium magnetic field $B_{\rm res}^{l}$ at $R_{i}$. The magnetic perturbation amplitude vector $\hat{\mathbf{B}}^{l}$ is modeled with a Gaussian radial envelope centered on $s_i$ with a width of $\Delta s_l=0.05$, scaled by a fractional amplitude $\delta B_{\rm frac}^l=3\times 10^{-3}$ of the equilibrium field $B_{\rm res}^l$, and oriented along the Cartesian unit vector $\hat{\mathbf{e}}^{l}=[0,0,1]$:
\begin{equation}
\hat{\mathbf{B}}^{l}
= \delta B_{\rm frac}^{l}\, B_{\rm res}^{l} 
\, e^{\!\left[-\frac{(s-s_i^{l})^2}{2\,\Delta s_l^2}\right]}
\hat{\mathbf{e}}^{l}    
\end{equation}  

The divergence-free condition ($\nabla\cdot\delta\mathbf{B}=0$) is satisfied by constructing the perpendicular amplitude vector $\hat{\mathbf{B}}_{\perp}^l$ as the projection of $\hat{\mathbf{B}}^l$ onto the plane orthogonal to the local wavevector $\mathbf{k}_l=\nabla\Phi_l$:
\begin{equation}
\hat{\mathbf{B}}_{\perp}^{l} = \hat{\mathbf{B}}^{l} - \frac{\mathbf{k}_l \cdot \hat{\mathbf{B}}_{l}}{\mathbf{k}_l \cdot \mathbf{k}_l}\,\mathbf{k}_l   
\end{equation}

Results show that for deuterium (see figures \ref{fig:D1}, \ref{fig:D0}), with different initial spin polarizations relax to a quasi-steady state after approximately 0.06\,ms. The depolarization is less pronounced when starting from the $m_I=+1$ state, which retains a final probability of 0.9260, compared to an initial $m_I=0$ state, which drops to a final probability of 0.8806. Another difference appears in how the other states are populated: when starting from $m_I=0$, the $m_I=+1$ and $m_I=-1$ states are populated equally. However, when starting from $m_I=+1$, the $m_I=0$ state is populated more than the $m_I=-1$ state. This asymmetry stems from the quantum mechanical selection rules for magnetic dipole transitions, which only allow spin transitions of $\Delta m_I = \pm 1$. As a result, a direct transition from $m_I = +1$ to $m_I = -1$ is forbidden, forcing the spin to pass through the intermediate $m_I = 0$ state---leading to its enhanced population while the $m_I = -1$ state remains suppressed.

On the other hand, results for tritium (see figure \ref{fig:T1}) show almost no depolarization. Tritium is lucky, despite its larger $g$-factor compared to deuterium, its spin-$\tfrac{1}{2}$ nature inherently makes it immune to depolarization. 

Overall, deuterium exhibits a measurable depolarization (up to $7.4\%$ in the $m_I=1$ state), whereas tritium remains nearly perfectly polarized in SPARC, under strong eigenmode perturbations over collisionless timescales.

Although bulk spin depolarization provides an overall estimate, it is also important to investigate the depolarization of individual ions in SPARC to better understand the underlying physics at the particle level. A resonant interaction between a wave and a moving spin-polarized nucleus occurs when the wave's frequency, $\omega$, satisfies the condition:
$$\omega = \omega_L + \ell \omega_c + k_{\parallel} v_{\parallel}$$ Here, $\omega_L$ is the Larmor frequency, $\omega_c$ is the cyclotron frequency, $k_{\parallel} v_{\parallel}$ is the Doppler shift (calculated from the wave's parallel wavenumber, $k_{\parallel}$, and the nucleus's parallel velocity, $v_{\parallel}$), and $\ell$ is the integer cyclotron harmonic number \cite{Kulsrud1986}. Spin-depolarization code reveals that depolarization is confined to a narrow, Doppler-shifted band around the fundamental Larmor resonance  ($\ell=0$), with negligible coupling to cyclotron harmonics. For this reason, depolarization is examined exclusively at the fundamental resonance. This is an encouraging result, as it means that only a single frequency band around the precession frequency can cause depolarization, while all higher harmonics remain benign. It is also important to note that in cases where multiple waves resonate with the spin polarized nucleus at a given location, the net magnetic perturbation amplitude, $\delta B$, is determined by the superposition of the waves. The outcome of this superposition depends on the individual wave phases, $\Phi_l$, as shown in equation \ref{eq: modes_sum}, which dictate the critical relative phases between them. Constructive interference between waves increases the effective perturbation amplitude and thereby enhances depolarization, whereas destructive interference reduces the net amplitude, potentially canceling the perturbations and preserving polarization. 

\begin{figure*}[t!]
    \centering
    \includegraphics[width=1\linewidth]{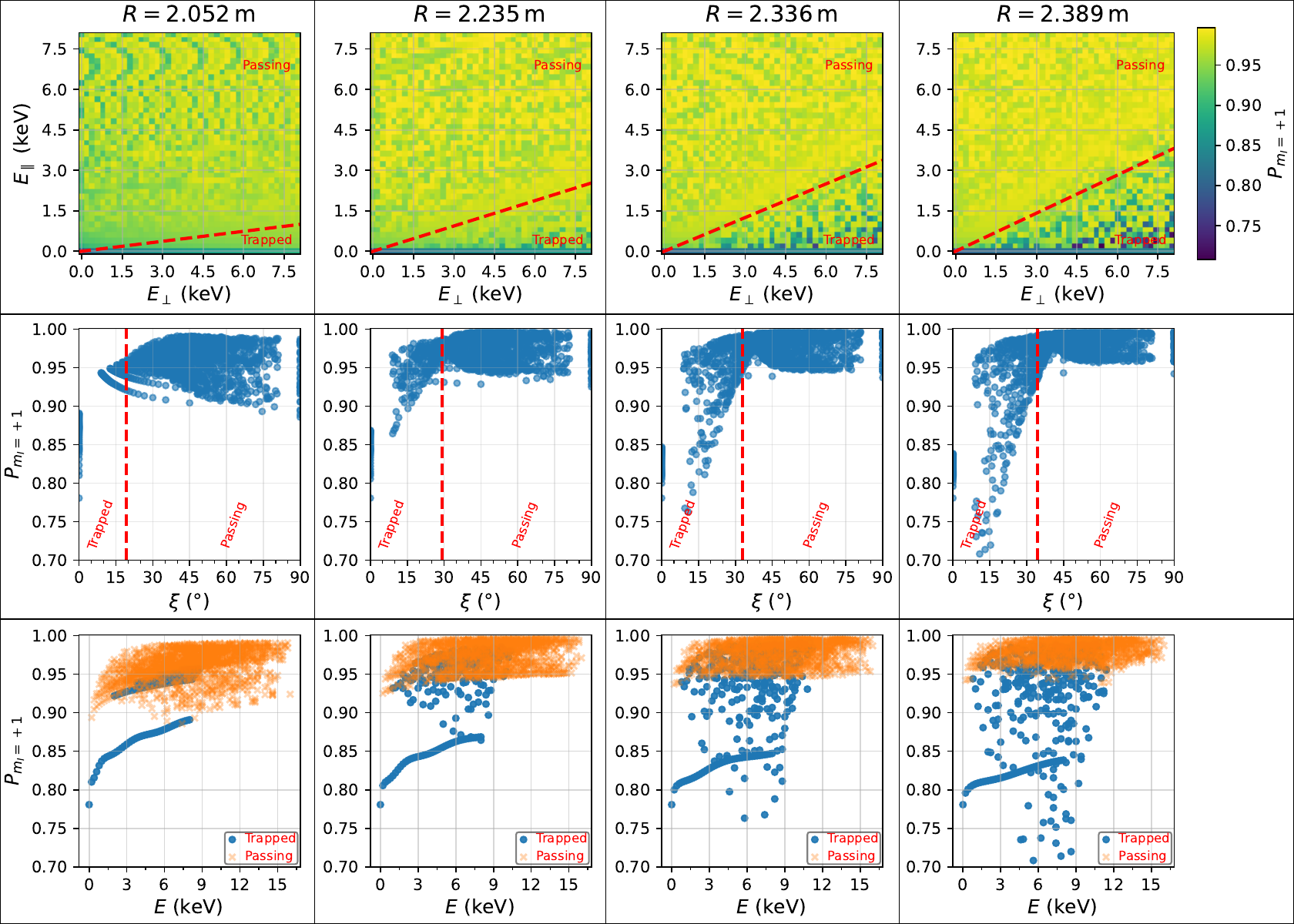}
    \caption{Mean temporal probability of a deuteron in the \(m_I=+1\) state, averaged over \(10^4\) cyclotron periods in SPARC, shown across \((E_{\parallel},E_{\perp})\) space (top row), pitch angle (middle row), and total energy (bottom row), with red dashed lines marking the trapped–passing boundary. Each column corresponds to a different major radius \(R_i\) on the outboard midplane where deuterons are initialized. The wave frequency is set to the local deuteron precession frequency, evaluated from the equilibrium magnetic field \(B_{\rm res}\) at \(R_i\), with perturbation \(\boldsymbol{\delta B} = (B_{\rm res}/1000\, {\hat e}_{\perp,1},\,0\,{\hat e}_{\perp,2},\,0\,{\hat e}_\parallel)\) and \(k=0\).}
    \label{fig:energy_grid}
\end{figure*}

\begin{figure}
    \centering
    \includegraphics[width=0.9\textwidth]{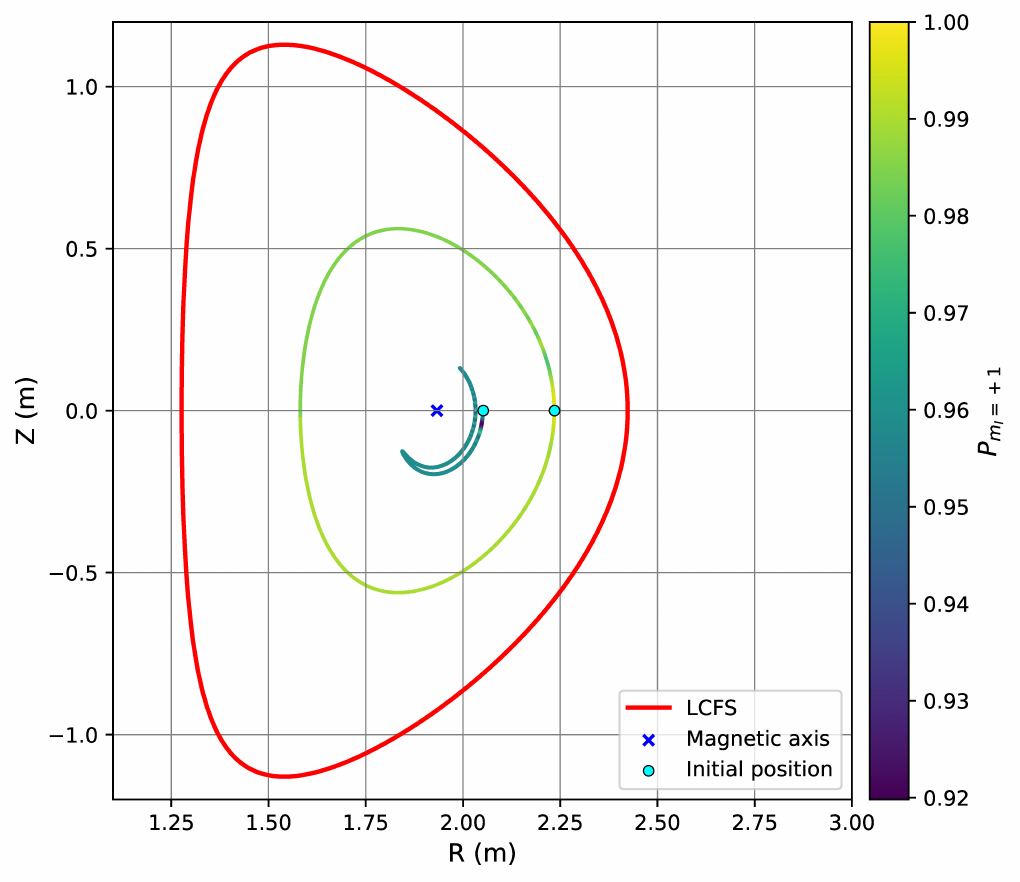}
    \caption{Trajectories of a trapped (banana) and passing (circular) deuteron in SPARC, showing the evolution over $10^4$ cyclotron periods for particles initialized in the $m_I=+1$ spin state. The color indicates the instantaneous probability of being in the $m_I=+1$ spin state. The trapped particle is launched at $R=2.052$ m with $E_{\parallel}=900$ eV, $E_{\perp}=8000$ eV, and the passing particle at $R=2.235$ m with $E_{\parallel}=3000$ eV, $E_{\perp}=3000$ eV.}
    \label{fig: particle_trajectory}
\end{figure}

To study the dependence of depolarization on particle energies, a deuteron in the spin-1 state is launched from the outboard mid plane at selected $R_{i}$ with prescribed parallel and perpendicular energies ($E_{\parallel}$, $E_{\perp}$) in SPARC. The wave frequency $\omega$ is set to the local deuteron precession frequency, evaluated from the equilibrium magnetic field $B_{\rm res}$ at $R_i$, with $\boldsymbol{\delta B} = [\,B_{\rm res}/1000,\,0,\,0\,]$ and $k=0$, so that Doppler-shift effects are neglected. Particles are classified as passing or trapped based on their pitch angle, \(\xi = \tan^{-1}\!\left(\sqrt{E_{\parallel}/E_{\perp}}\right)\), with passing particles satisfying \(\tan^{-1}(\sqrt{2\epsilon}) < \xi < \pi/2\) and trapped particles satisfying \(0 < \xi < \tan^{-1}(\sqrt{2\epsilon})\). Here, \(\epsilon = r/R\) is the inverse aspect ratio of the flux surface at the particle’s initial position, with \(r\) the minor radius and \(R\) the major radius. Each simulation is run over \(10^4\) cyclotron periods (collisionless timescales), and the resulting mean temporal probability is evaluated as shown in figure \ref{fig:energy_grid}. In addition, figure \ref{fig: particle_trajectory}, also shows example trajectories for a trapped particle (inner banana orbit) and a passing particle (outer circular orbit), where the color indicates the instantaneous probability of occupying the $m_I=+1$ spin state, obtained using the same simulation setup described in this paragraph.

The analysis across four flux surfaces in SPARC (see figure \ref{fig:energy_grid}) shows that depolarization is most pronounced in the trapped region of phase space (low $E_{\parallel}$, high $E_{\perp}$) and becomes more significant at larger radii, where the trapped-particle fraction increases. Passing particles show the strongest and most scattered depolarization near the core, but as the radius increases, the spread decreases and the probabilities cluster at higher values. Overall, orbit topology (trapped vs. passing) rather than total energy is the dominant factor for depolarization, and it is favorable that the trapped fraction remains small in the core where fusion reactions predominantly occur; however, a fraction of passing particles near the core are more prone to depolarization. 

These results can be explained by noting that the cumulative spin depolarization depends on the relative time particles spend in resonance versus off resonance: extended time in resonance enhances depolarization, while longer recovery periods off resonance mitigate it and help restore polarization. Differences in orbit geometry between trapped and passing particles affect the relative time spent in and out of resonance, making trapped particles generally more prone to depolarization. A fraction of passing particles near the core, confined to smaller-radius flux surfaces, exhibit stronger depolarization, whereas those on larger-radius flux surfaces depolarize less, as their extended trajectories reduce the relative time spent in resonance and increase time in non-resonant regions.

\section{Discussion} \label{sec:discussion}

In this work we have shown a comprehensive numerical investigation of spin depolarization due to waves in plasma with frequencies comparable to the ion-cyclotron frequency. In particular, we focused on calculating depolarization in the SPARC tokamak at different radial locations and across deuterium and tritium energies. Overall, we have found the depolarization is weak over tens of thousands of ion cyclotron periods. While further work is required to study the evolution over longer time periods, these results suggest that spin-polarized deuterium and tritium could persist for sufficiently long enough to increase power density in magnetic confinement fusion machines.

To conduct this study, we developed a tool capable of self-consistently integrating the position, velocity, and particle spin state of ions in a superposition of a static uniform magnetic field and an arbitrary number of spatially and temporally oscillating magnetic fields. Using this framework, we identified normal modes supported by an infinite homogeneous D–T plasma with the potential to depolarize deuterium based on their left-handedness, frequency, and perpendicular component amplitude. Our analysis shows that waves on the ion–ion hybrid branch may drive deuterium depolarization, while parallel-propagating waves on the fast Alfvén branch may similarly affect tritium.

We studied the role of alpha-particle birth distributions arising from spin-polarized fuel schemes, evaluating their ability to destabilize waves in a manner analogous to calculations performed for ion cyclotron emission. These distributions are found to emit waves obliquely to the magnetic field and only 20\% larger than alpha-particle shell distributions generated from unpolarized fuel. Fully nonlinear electromagnetic PIC simulations revealed the handedness and relative strength of obliquely propagating Alfvén waves and showed that they represent the dominant risk to spin polarization. Compared to zero parallel wavenumber, a small but finite value amplifies relative magnetic fluctuation amplitudes by two to three orders of magnitude and drives strong left-handed polarization near the D and T precession frequencies. Notably, the choice of fuel polarization scheme does not alter wave magnitude or polarization, while increasing the alpha fraction raises the mean magnetic fluctuation amplitude without changing its polarization characteristics.

Monte Carlo integration was used to evaluate the spatio-temporal expectation value of the deuterium fuel polarization state in SPARC-like equilibria. These calculations indicate that depolarization remains modest even on collisional timescales, depending on the spatial structure of the magnetic perturbations. Tritium, by contrast, remains completely resistant to depolarization because of its spin-1/2 character, underscoring the robustness of spin-1/2 fuels for polarized fusion applications. Finally, we highlight that high magnetic field strength could provide an intrinsic advantage for preserving spin alignment. Since the efficacy of depolarization depends on the relative perturbation $\delta B/B$ rather than the absolute amplitude $\delta B$, operation at higher $B$ naturally suppresses this ratio if $\delta B$ remains constant, offering an important mechanism to sustain polarization in next-generation high-field devices.

\section{Acknowledgments}

We acknowledge funding from Stellar Energy Foundation. We are grateful for conversations with R. W. Engels, W. W. Heidbrink, A. Rutkowski, and J. A. Schwartz. We acknowledge the computational resources of the Flux HPC cluster at the Princeton Plasma Physics Laboratory used in this research work.



\appendix

\section{ITER Results} \label{sec:ITER}
\begin{figure*}[t]
    \centering
    \includegraphics[width=0.99\linewidth]{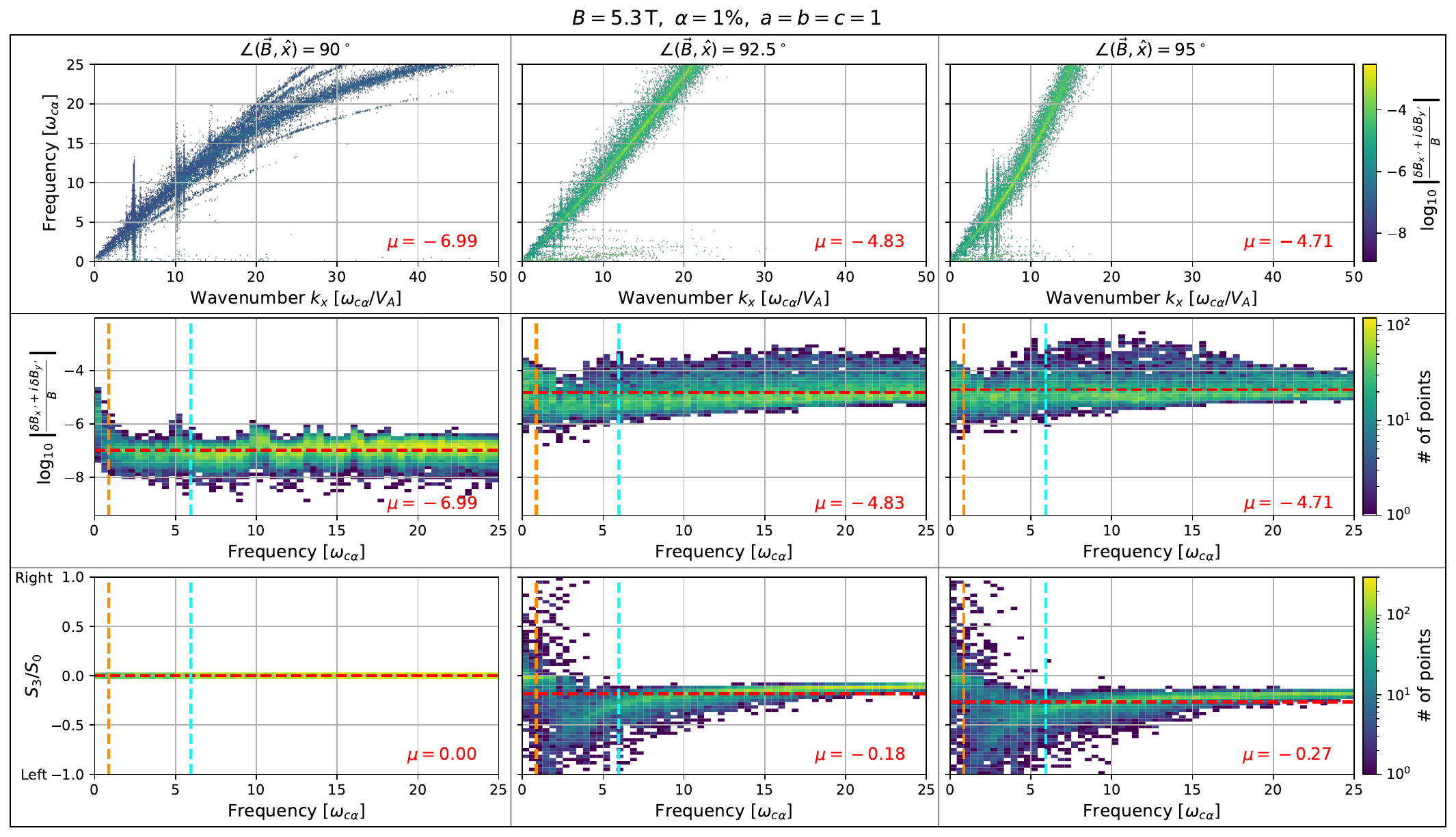}
    \caption{EPOCH simulation results for ITER at three magnetic field orientations relative to the simulation domain ($\angle(\vec{B},\hat{x}) = 90^\circ,\,92.5^\circ,\,95^\circ$, shown column-wise), with constant alpha fraction ($\alpha = 1\%$) and fuel polarization ($a=b=c=1$).}
    \label{fig:ITER_angle}
\end{figure*}

\begin{figure*}[t]
    \centering
    \includegraphics[width=0.99\linewidth]{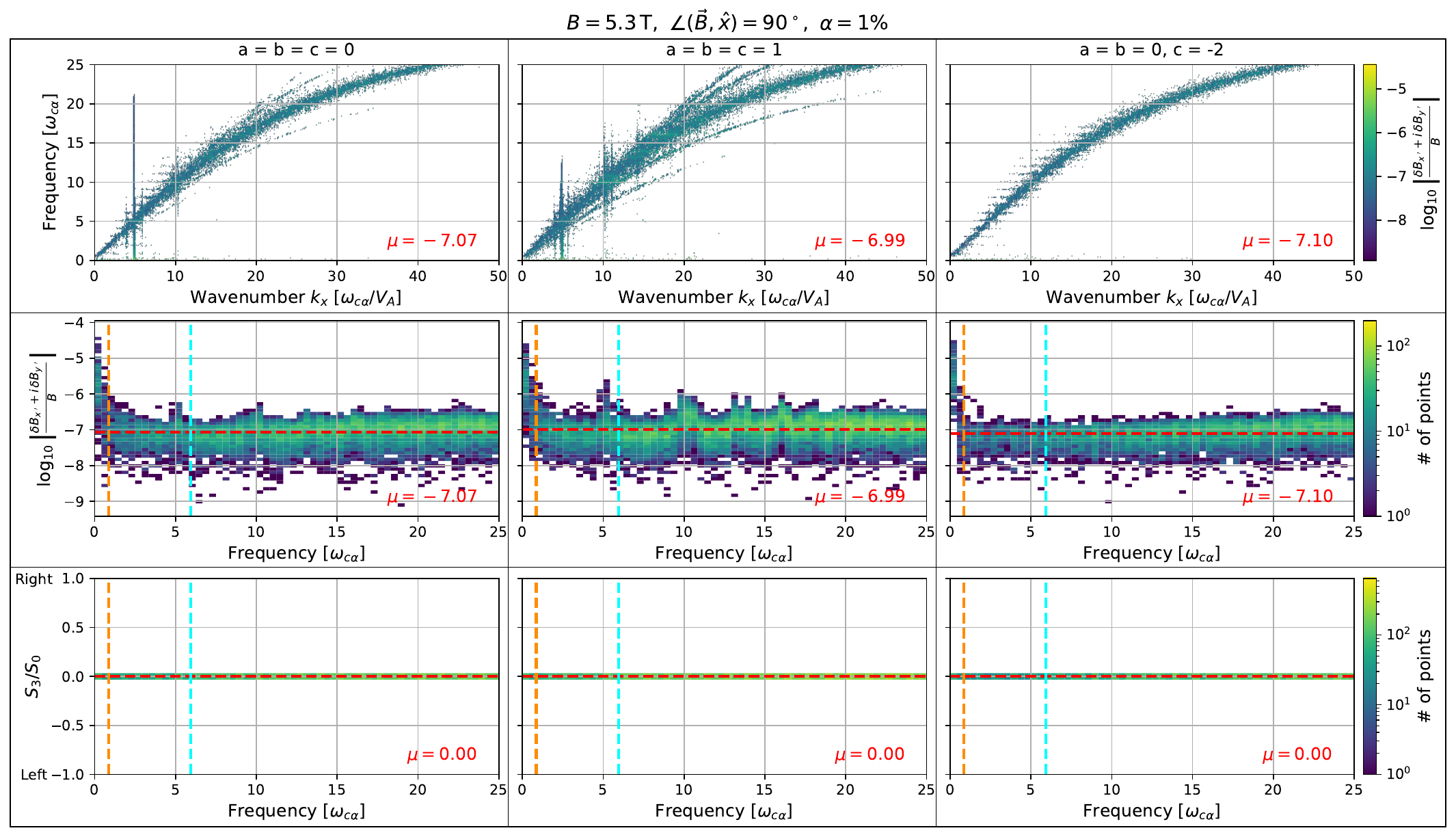}
    \caption{EPOCH simulation results for ITER at three fuel polarization configurations ($a=b=c=0$, $a=b=c=1$, and $a=b=0,\,c=-2$, shown column-wise), with constant magnetic field orientation $\angle(\vec{B},\hat{x}) = 90^\circ$, and alpha fraction $\alpha = 1\%$.}
    \label{fig:ITER_fuel}
\end{figure*}

\begin{figure*}[t]
    \centering
    \includegraphics[width=0.99\linewidth]{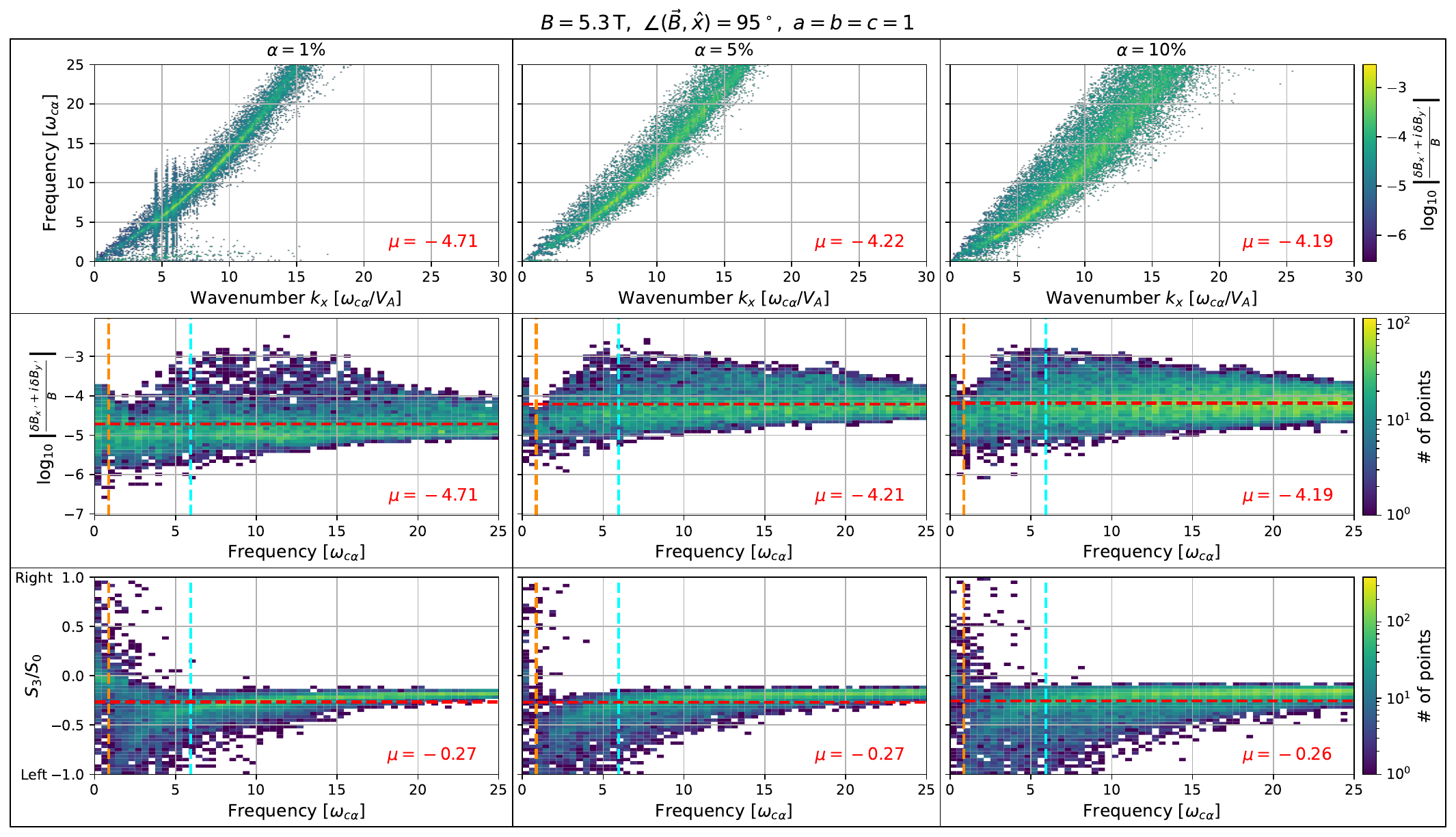}
    \caption{EPOCH simulation results for ITER at three alpha fractions ($\alpha = 1\%,\,5\%,\,10\%$, shown column-wise), with constant magnetic field orientation $\angle(\vec{B},\hat{x}) = 95^\circ$, and fuel polarization $a=b=c=1$.}
    \label{fig:ITER_alpha}
\end{figure*}

The PIC simulation results for ITER (figures \ref{fig:ITER_angle}, \ref{fig:ITER_fuel}, and \ref{fig:ITER_alpha}) reproduce the SPARC trends: oblique magnetic field angles (small finite $k_{\parallel}$) amplify magnetic fluctuations and drive a transition toward left-handed polarization. The fluctuation amplitude increases with the fast-alpha fraction, yet is largely insensitive to the chosen fuel-polarization scheme. The Stokes ratio $S_3/S_0$ is independent of both the alpha fraction and the fuel polarization.

\bibliography{bibliography} %

\end{document}